\documentclass[twocolumn,superscriptaddress,showpacs,amsfonts]{revtex4-1}
\pdfoutput=1

\usepackage{epsfig}
\usepackage{graphicx}
\usepackage{amsmath}
\usepackage{amssymb}
\usepackage{color}
\usepackage{bm}

\begin{document} 
\title{Topological Gapless Phases in Non-Symmorphic Antiferromagnets} 
\author{Wojciech Brzezicki}
\affiliation{CNR-SPIN and Dipartimento di Fisica ``E. R. Caianiello'',
Universit\`a di Salerno, I-84084 Fisciano (Salerno), Italy}
%
\author{Mario Cuoco}
\affiliation{CNR-SPIN and Dipartimento di Fisica ``E. R. Caianiello'',
Universit\`a di Salerno, I-84084 Fisciano (Salerno), Italy}
\begin{abstract}
Topologically protected fermionic quasiparticles occur in metals with band degeneracy as a consequence of band structure topology. 
Here we unveil topological semimetal and metal phases in a variety of non-symmorphic collinear antiferromagnets with glide reflection 
symmetry, a combination of mirror and half-lattice translation. We find gapless phases with  
Dirac points having multiple symmetry-protection as well as electronic structures with triple and quadruple band-crossing points. 
Glide semimetal is shown to be converted into a topological phase 
with non-trivial $\mathbb{Z}_2$ topological charges at the Dirac points due to inversion and time-inversion symmetry combination. 
More striking is the emergence of a 
hidden non-unitary relation between the states in the glide sectors that provide a general mechanism to get multiple band touching points. 
The split Fermi points uncover a $\mathbb{Z}_2$ protection that drives the changeover of the multiple-degenerate gapless phase into a topological 
metal built from their connection through distinct Fermi lines.  
Besides a new perspective of ordered states in complex materials, our findings indicate that 
novel topological gapless phases and edge states may occur in a wide class of magnetic systems. 
\end{abstract}

%

\maketitle
\noindent 
\section{Introduction \label{sec:intro}} 
Topological materials have become the focus of intense research in the last years~\cite{Hasan2010,Qi2011,Chiu2016,Weng2015a} not only
for the perspective of new physical phenomena with potential technological applications, but also for being a test bed for fundamental concepts of physics theories. 
Along this line, recent efforts led to the theoretical 
prediction \cite{Kane2005,Bernevig2006,Moore2007,Fu2007} and experimental realization \cite{Konig2007,Hsieh2008,Xia2009a} of topological 
insulators in materials with strong spin-orbit coupling (SOC).
One of the hallmarks of TIs is the existence of protected gapless edge states, which are due to a non-trivial
topology of the bulk band structure. Such manifestation of topological order however is not limited to insulators 
as electronic structures with gapless topological modes have been predicted~\cite{Volovik2003,Volovik2007,Wan2011,Heikkila2011,Yang2011,Burkov2011,Krempa2012,Xu2011,Halasz2012}
and their relevance further boosted by the discovery of novel materials~\cite{Weng2015b,Huang2015,Xu2015a,Lv2015a,Lv2015b,Xu2015b} 
with non-trivial band crossing points in the momentum space and robust edge states. 

\begin{figure}[t]
\includegraphics[clip,width=1\columnwidth]{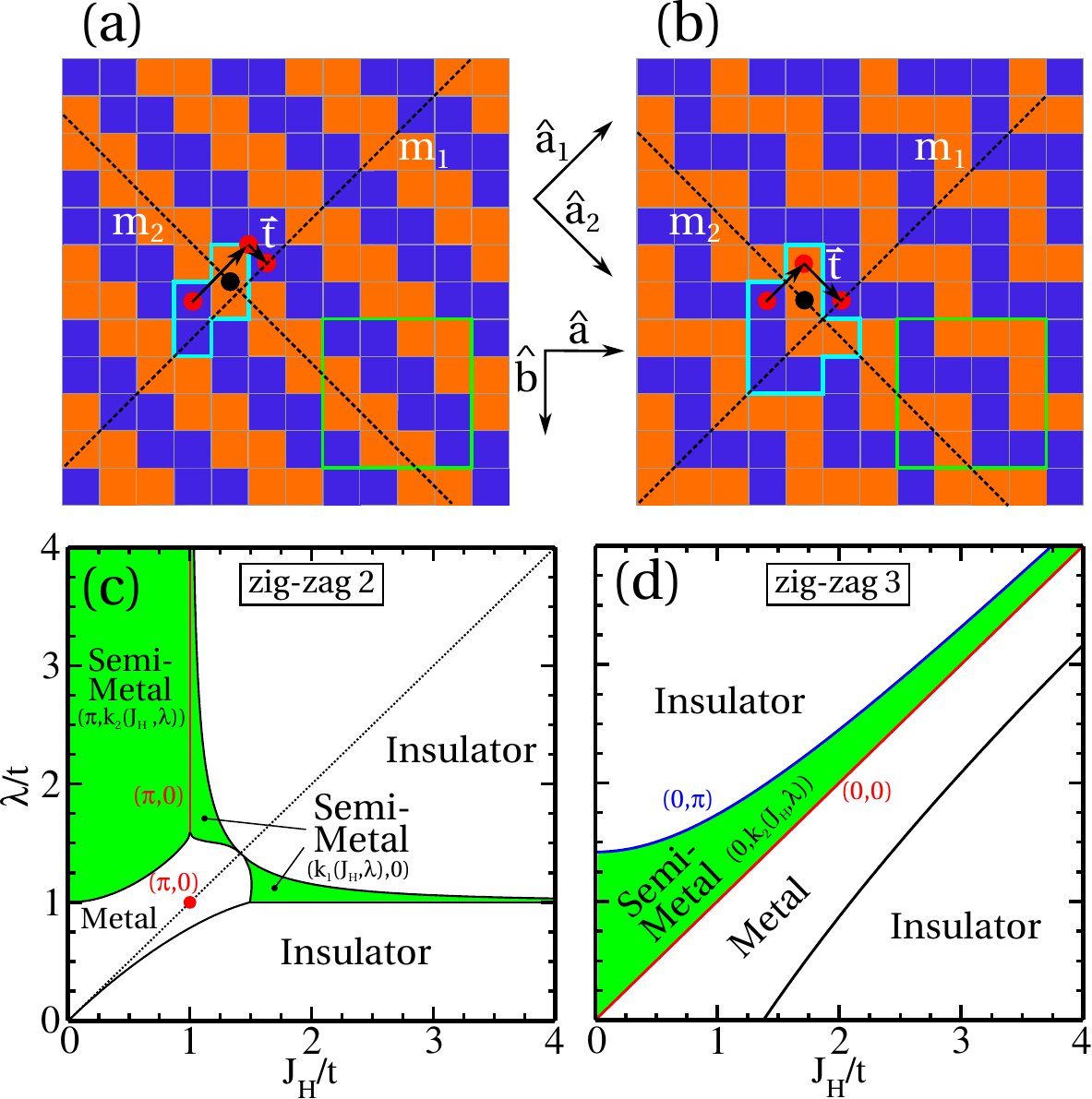}
\protect\caption{
(Color online) Zig-zag spin patterns with (a) length $L_{z}=2$ ($z2$) and (b) 
$L_{z}=3$ ($z3$). Orange and blue squares indicate spin-up and spin-down orientation. 
The unit cell is marked by a thick
blue frame while $\hat{a}_{1,2}$ are the translation directions. The
square unit cells are marked with green frames. Dashed lines indicate the
normal mirror planes $m_{1}$ and $m_{2}$, with $m_{2}$ being related to the
the gliding symmetry. The glide transformations is sketched via
the red dots and arrows; the dot is subjected to the reflection $m_{2}$
then it is translated by a vector $\vec{t}$ parallel to the mirror
plane $m_{2}$. For $z2$ these two steps separately
do not reproduce the original lattice. The black dots are the inversion
centers. Phase diagrams for (c) $z2$ at $3/4$ filling and
(d) $z3$ at half filling. Topological semimetal phases are marked
with color (green area) and the position of Dirac points in the Brillouin zone is reported in parenthesis.
Red and blue lines in (c) and (d) indicate configurations where the Dirac points are in high symmetry positions.
\label{fig:patt_PDs}}
\end{figure}

Among various kinds of topological matter, correlated materials~\cite{Imada1998,Tokura2000} with strong spin-orbital-charge entanglement~\cite{Oles2012,Vojta2009,Brz2015,Brz16}, 
e.g. transition-metal oxides, represent a unique platform to explore topological effects
combined to a large variety of intriguing collective properties emerging from electron-electron interaction, as superconductivity, magnetism, 
magnetoelectricity and Mott insulating phases.
In these systems, complex magnetic orders generally arise from competing ferromagnetic (FM) and antiferromagnetic (AF) correlations with
a frustrated localized-itinerant nature
and a strong dependence on the 
orbital character of the transition metal $d$-shells.
Magnetic patterns constructed by antiferromagnetically coupled zig-zag 
FM chains (Figs. \ref{fig:patt_PDs}(a),(b)) 
are one generic manifestation of such competing effects and often occur in the class of 
correlated materials as demonstrated in manganites~\cite{Tokura2006,Munoz2001,Dong2015}, ruthenates~\cite{Ortmann2013,Mathieu2005,Hossain2012,Mesa2012}, 
dichalcogenides~\cite{Xia2009b,Chen2009,Fobes2014}, iridates~\cite{Ye2012,Choi2012}, nickelates~\cite{Munoz1994,Alonso1999, Diaz2001, Zhou2005}, etc. 
A relevant mark of zig-zag patterns is the symmetry under non-symmorphic (NS) transformations that combine point group 
operations with translation that are a fraction of a Bravais lattice vector~\cite{Bradley1972}. 
Recently, NS groups have been recognized as a new source of topological symmetry protection both in gapped~\cite{Mong2010,Liu2014,Fang2015,Shiozaki2015,Dong2016,Wang2016,Varjas2015,Sahoo2015,Lu2016,Shiozaki2016a,Kobayashi2016,Alexandradinata2016,Chang2016,Wang2016Nat} and gapless~\cite{Parameswaran2013,Watanabe2015a,Young2015,Watanabe2016,Liang2016,Venderbos2016,Yang2016,Muechler2016,Zhan2016,Bzdusek2016,Wieder2016,Bradlyn2016,Chen2016,Wieder2016a,Pixley2016} systems. Hence, given the wide range of physical phenomena
in both topological and correlated materials, the identification of novel topological phases in the presence of non trivial orderings and their material realizations represent a fundamental challenge in the condensed matter area.
 
In this paper, we unveil topological semimetal (SM) and metal phases in a variety of non-symmorphic 
collinear antiferromagnets with glide reflection 
symmetry. The emergent topological gapless states can exhibit   
Dirac points (DPs) with multiple symmetry-protection as well as three- and four-fold degenerate DPs. 
Besides non-symmorphic symmetry protection, 
we demonstrate that combination of inversion and time or particle-hole symmetry can lead to 
non-trivial $\mathbb{Z}_2$ topological charges at the DPs, thus 
building up a robust topological gapless phase. 
More striking is the occurrence of a 
hidden symmetry-like relation between the states in the two glide sectors that provide a 
general mechanism to stick DPs or generate 
multiple band touching points in the glide plane (GP). 
We show how the splitting of the multiple degenerate Fermi points (FPs)
drives the transition into a topological metal. Due to a $\mathbb{Z}_2$ protection of the DPs in the glide plane, 
a topological metal arises with multiple Fermi pockets resulting from the 
Fermi lines connecting the DPs themselves. 

The paper is organized as follows. In Sec. II we present the model Hamiltonian employed to describe the zigzag antiferromagnets and the related symmetries. Sec. III is devoted to discuss the 
phase diagram and the most relevant topological phases emerging among the obtained electronic structures. 
In the Sec. IV, we provide the concluding remarks. The various technical details including some representative cases of
energy spectra for the zig-zag antiferromagnetic phases are given in the Appendices \ref{sec:struc}-\ref{sec:topo_inv}.

\section{Model and symmetries \label{sec:model}}
We consider an effective orbital-directional double-exchange model describing itinerant electrons (e.g. t$_{2g}$ or $p$ bands)
in the presence of an anisotropic SOC, as due to tetragonal crystal field splitting, 
and Hund coupled to localized spin moments forming zig-zag pattern with characteristic length $L_z\ge 2$.  
The model Hamiltonian is given by
\begin{eqnarray}
{\cal H} & = & \sum_{i,\sigma}\sum_{{\alpha,\beta=a,b\atop {\hat{\gamma}}={\hat{a} ,\hat{b}}}}t_{\gamma,\alpha \beta} \left(d_{i,\alpha\sigma}^{\dagger}d_{i+\hat{\gamma},\beta\sigma}+h.c.\right)\nonumber \\
 & - & J_{H}\!\!\!\sum_{i,\alpha=a,b}\!\! s^z_{i\alpha}\cdot S^z_{i} +
\lambda\sum_{i} S_{i}^{z} l_{i}^{z} \label{eq:Ham}  \nonumber \,,
\end{eqnarray}
where $d^{\dagger}_{i,\alpha\sigma}$ is the electron creation operator at the site $i$ with spin $\sigma$ for the orbital
$\alpha$, $(a,b,c)$ are the $(yz,xz,xy)$ orbitals which
are perpendicular to the corresponding
bond direction, with $\hat{a}$, $\hat{b}$, and $\hat{c}$ being the unit vectors along the lattice symmetry directions.
$\alpha$, ${\bf s}_{i\alpha}\!=\! d_{i,\mu,\alpha}^{\dagger}\sigma^z_{\!\alpha,\beta}d_{i,\mu,\beta}$
and $S^z_{i}=\pm 1$ denote the spins for the $d_{xz/yz}$ and
$d_{xy}$ orbitals, respectively.  $J_{H}$ stands for the Hund coupling, while $\lambda$ is 
the SOC for the projected subspace of $(a,b)$ orbitals, with
$l_{i}^{z}=i\left(d_{i,a,\sigma}^{\dagger}d_{i,b,\sigma}-d_{i,b,\sigma}^{\dagger}d_{i,a,\sigma}\right)$ the $z$-component of the local 
angular momentum.
$t_{\hat{\gamma},\alpha \beta}$ is the nearest-neighbor hopping amplitude between $\alpha$ and $\beta$
for the bond along the $\hat{\gamma}$ direction. 
We assume cubic symmetric hopping amplitudes, i.e., $t_{\hat{a},bb}=t_{\hat{b},aa}=-t$ with $t$ as energy scale unit and $t_{\hat{\gamma},ab}=0$. 
The AF states are collinear and the spin $z$-projection is a conserved quantity due to the anisotropic SOC.
The exact matrix forms of the $k$-space Hamiltonians ${\cal H}_{\vec{k}}$ for $L_z=2,3$ are given in the Appendices \ref{sec:struc} and \ref{sec:hams=000026syms_gaug}.

While the Hamiltonian resembles the double-exchange model widely applied in the context of manganese oxides, it contains extra microscopic ingredients as orbital directionality and spin-orbit coupling that contribute to give unique features in the phase diagram and the electronic spectra.  For instance, zig-zag states have been demonstrated to be among the energetically most favorable configurations in a large range of doping concentration for the case of $d_{xz/yz}$ bands~\cite{Brzezicki2015} and to be relevant for providing novel scenarios and mechanisms to understand the occurrence of zig-zag spin patterns with length $L_{z}=2$ (i.e. $z2$) in Mn doped bilayer ruthenates Sr$_3$Ru$_2$O$_7$ and in other similar hybrid oxides. 
Furthermore, it can be considered as an effective low energy description of correlated $t_{2g}$ electrons
in transition metal oxides with orbital selective localized and itinerant bands such as to yield a double exchange model. Indeed,
as a consequence of the atomic Coulomb interaction in multi-orbital systems, an orbital selective Mott transition can occur and lead to electronic localization in a subgroup of the $t_{2g}$ bands. Such reduction from a multi-orbital correlated system to an effective double-exchange has been generally addressed and demonstrated to be applicable in the context of orbital-selective-Mott physics \cite{orbsel-PRL05,orbsel-PRL14}.

The symmetry properties of the model Hamiltonian include transformations that act on the internal (e.g. spin, orbital, charge) and spatial degrees of freedom, or combine them, including the possibility of having groups with non-primitive lattice vector translational (e.g. non-symmorphic groups). 
Concerning the internal symmetries, the model exhibits only time reversal invariance, ${\cal T}$ with ${\cal T}{^2}=1$ and ${\cal T}^{\dagger} {\cal H}_{\vec{k}} {\cal T}= {\cal H}^T_{-\vec{k}}$. 

For the spatial symmetries, electrons move on a square lattice and within a magnetic pattern due to a broken symmetry ground-state. 
The model Hamiltonian can exhibit invariance under a reflection plane $m_{1}$ and 
has inversion centers (see Figs.
\ref{fig:patt_PDs}(a)-(b)) associated to the reflection and inversion operators, $\cal R$ and $\cal I$ respectively.
Then, in two dimensions it is known that there are other relevant symmetry operations belonging to the non-symmorphic groups which include 
screw axis, glide mirror lines, and glide mirror planes in conjunction with 
the translation~\cite{Bradley1972}. 
The zig-zag antiferromagnet (AFM) is invariant under a NS glide transformation ${\cal R}^{t}$
which is constructed by the product of a reflection with respect to the $m_{2}$ plane and
a translation $\vec{t}\equiv\vec{a}_{2}/2$ in the $\hat{a}_2$ direction along the zig-zag chain (Figs.
\ref{fig:patt_PDs}(a)-(b)). 
Due to the multi-orbital character of the model Hamiltonian, both ${\cal R}$ and ${\cal R}^{t}$ 
include spatial and orbital transformations. Indeed, the mirror reflections interchange
the $\hat{a}$ and $\hat{b}$ lattice directions and consequently the 
orbitals, see Appendix \ref{sec:spa_sym}. 
${\cal R}^{t}$ is also intrinsically $k$-dependent as it is not possible to find a unit
cell that maps onto itself under the glide transformation. 
By a proper choice of the unit cell or a suitable $k$-dependent 
transformation of ${\cal H}_{\vec{k}}$, as thoroughly discussed and demonstrated in the Appendix \ref{sec:gaug_refl2}, one can show that the glide operator depends only on $k_{2}$ as ${\cal R}_{k_{2}}^{t}$.
We select a basis that makes ${\cal R}_{k_{2}}^{t}$ to have eigenvalues
$g_{\pm}=\pm1$ while the eigenvectors carry the $k_{2}$-dependence. 
As expected from the half-lattice translation of NS symmetry, the glide eigenstates have a doubled 
period in the momentum space. It is worth pointing out that ordinary reflection ${\cal R}$ 
is also $k$-dependent for the present choice of the unit cell (see Fig. \ref{fig:cell} in the Appendix for
a detailed view and description of the unit cell)
but it can be made completely $k$-independent by a suitable gauge transformation (as shown in 
Appendix \ref{sec:gaug_refl}) without affecting the periodicity of ${\cal H}_{\vec{k}}$.

At half filling, the model Hamiltonian exhibits a non-symmorphic like chirality operator ${\cal S}_{k_1}$ (and then also a particle-hole ${\cal C}_{k_1}$), as it carries an intrinsic $k_1$-dependence \cite{Shiozaki2016a} and acts to yield  
${\cal S}_{k_1}^{\dagger} {\cal H}_{\vec{k}} {\cal S}_{k_1}= -{\cal H}_{\vec{k}}$. The details of the structure and the properties are explicitly reported and discussed in the Appendix \ref{sec:non-spa_sym}.
As for the chirality, also the inversion ${\cal I}$ 
symmetry carries an intrinsic $k$-dependence which is tightly 
linked to the structure of the zig-zag magnetic pattern and the non-symmorphic 
glide symmetry. 
A peculiar feature of the symmetry properties is that their $k$-dependence can be removed 
only by allowing a period elongation for the eigenstate of the Hamiltonian in the Brillouin zone, 
as explicitly demonstrated in the Appendix \ref{sec:gaug_all}.
The reason for that is ascribable to the lack of a unit cell that would map onto itself under the
action of these operators. 
Concerning the relations between the
symmetry operators, all spatial symmetries commute between each other and with
time reversal while reflection $\cal R$ commutes also with chirality. Glide
and inversion commutes/anticommutes with chirality (thus also $\cal C$)
for odd/even $L_z$, see Appendices \ref{sec:commu}. 
Combined spatial-non-spatial symmetries can also be 
relevant for the character of the electronic structure and thus for completeness their properties are explicitly reported 
in the Appendix \ref{sec:combi}.

Finally, concerning the symmetry aspects of the model Hamiltonian, we also 
mention that there occur special symmetries that act only in the 
parameter space. Indeed, 
one can identify reflections in the parameter space of the Hund and spin-orbit couplings $(J_H,\lambda)$ which can be 
expressed by an SU($2$) algebra through their generators $\cal X$ and $\cal Y$.
The explicit form of the parameters space reflection symmetry is reported in the Appendix \ref{sec:super}.


\section{Results \label{sec:res}}

Concerning the electronic behavior of zig-zag AFM, our focus is on $z2$ and $z3$ magnetic patterns
which are those more relevant for the materials perspective as discussed in the introduction. 
A generic feature of the electronic phase diagram is that both SOC and Hund interaction are able to yield an insulating state that can be generally ascribed to the formation of almost disconnected orbital molecules developing within the zig-zag path due to the orbital directionality of the $d_{xz/yz}$ bands \cite{Brzezicki2015}. Then, the relative ratio of the SOC and Hund coupling can drive a series of (semi)metal-insulator transitions where non-trivial gapless phases are robust to variation of the microscopic parameters or occur
in between the insulating states as demonstrated for two representative electron 
densities in Figs. \ref{fig:patt_PDs} (c),(d). Another relevant aspect of the electronic phase diagram is that the semimetal states can have Dirac points in the normal mirror, in the glide planes, or in a generic position of the Brillouin zone, thus indicating that both non-spatial and spatial symmetries can be crucial in protecting the Fermi surface. 
Moreover, along the diagonal of the phase diagram a semimetal phase is obtained with Dirac points always lying in one of the glide plane, i.e. at $k_1=0$. Such finding is specific of the model Hamiltonian and is a consequence of a symmetry in the parameter space that interchanges $J_H$ with $\lambda$ or $J_H$ with $-\lambda$ in ${\cal H}_{\vec{k}}$,
as also described in more details in Appendix \ref{sec:super}. 

%
%
\begin{figure*}[t]
\includegraphics[clip,width=1\textwidth]{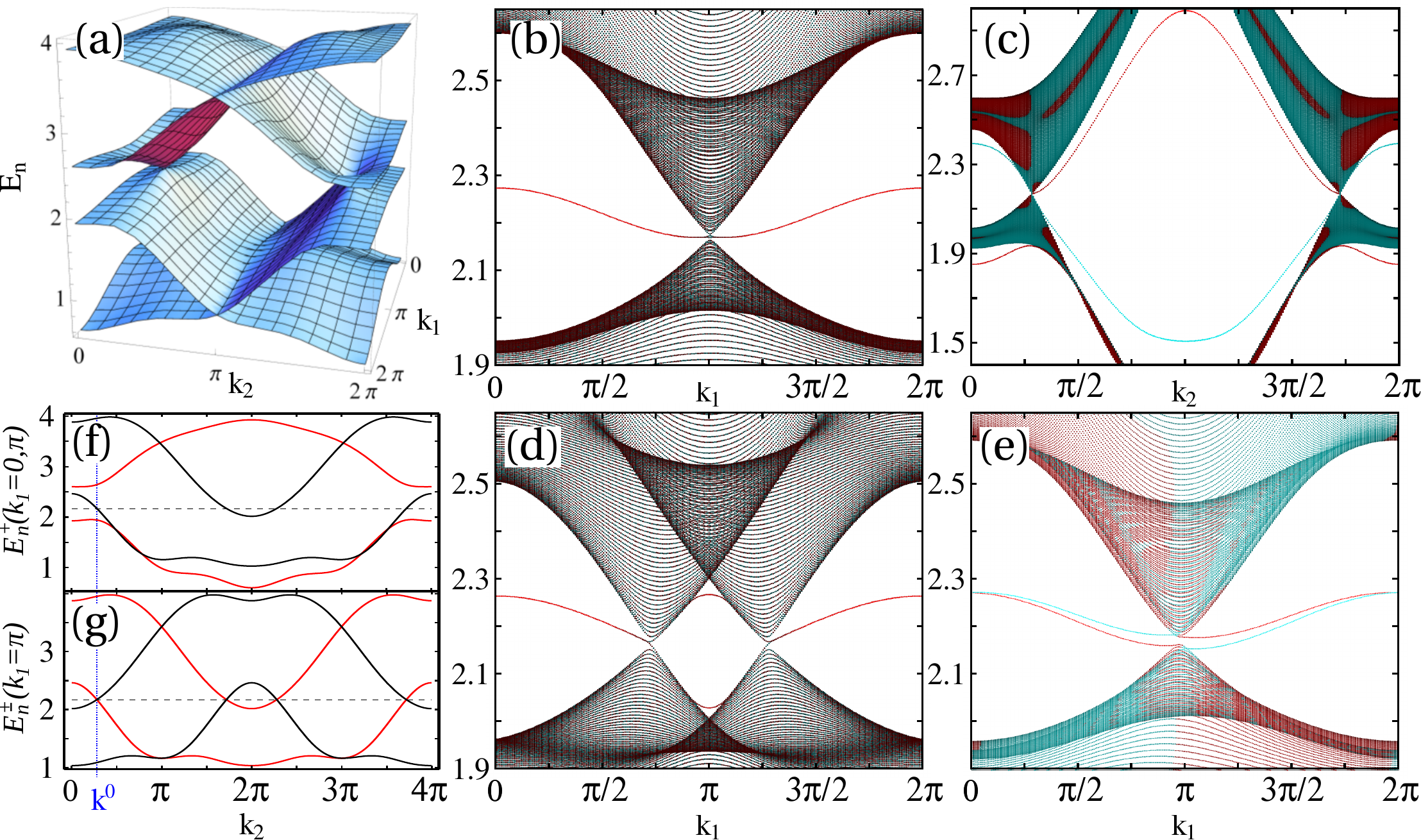}
\protect\caption{
Electronic spectra for $z2$ AF at $3/4$ filling. (a) Bulk energy spectrum above zero energy. 
(b)-(e) one-dimensional spectra and edge states for a slab geometry with open boundary. Spectra for a slab configuration with open boundary 
in the (b) $\hat{a}_1$ and (c) $\hat{a}_2$ directions, with $k_{2}$ and $k_{1}$ being the 
momentum parallel to the edge, respectively. Electronic spectra with (d) open boundary along $\hat{a}_2$, broken glide and reflection symmetries with
inversion invariance, and (e) with only broken time-reversal. (f) $g_{+}$ glide symmetric bands at $k_{1}=0$ (red) and $k_{1}=\pi$ (black) as functions of 
$k_{2}$. 
(g) Band structure in the glide plane at $k_{1}=\pi$, with glide eigenvalues $+1$
(red) and $-1$ (black). Dashed line sets the Fermi level.
Color map of the spectra in (b)-(e) with high (low) brightness indicates a large (small) probability of the electronic states to be localized
on the left (red) or right (blue) boundary. 
\label{fig:z2_sm}}
\end{figure*}
%


For the topological properties of the resulting electronic phases, we start discussing the insulating phases. According to the character of the internal symmetries of the model Hamiltonian (i.e. ${\cal T}^2=1$) the system is in the AI class of the ten-fold Altland-Zirnbauer classification table \cite{Altland}. For this case, in two-dimensions the fully gapped states cannot have any topological protection due to the non-spatial symmetries. 
On the other hand, for the gapless phases (Fermi lines or points) non-trivial topological invariants are allowed in the AI class. 

The presence of symmetries that act nonlocally in position space can in principle expand the possibilities of having non-trivial topological properties both for the insulating and the gapless phases. For the AI class it turns out that the insulating phase in two-dimension are always trivial both in the presence of normal mirror symmetry \cite{Sch14} or non-symmorphic transformations \cite{Shiozaki2016a}.
Hence, the main aim is to investigate the eventual topological character of the gapless states. 

We start by considering the $z2$ AFM and we focus on the DPs in the glide plane because our primary interest is to address the role of the non-symmorphic glide symmetry. 
We will see that not only the glide, but also the inversion, and combination of inversion with time-reversal combination are relevant symmetries for protecting the semimetal phase (Fig.\ref{fig:z2_sm}).

To set the stage, let us start discussing the band structure in the GP at $k_{1}=\pi$ (Fig. \ref{fig:z2_sm}(g)) which is representative of the electronic phase at large spin-orbit coupling in the diagram of Fig. 1(c) . Since the glide ${\cal R}_{k_{2}}^{t}$ commutes
with ${\cal H}_{\vec{k}}$, the electronic states can be labeled by $g_{\pm}$ glide eigenvalues.
The DPs occur at the crossing of the bands with opposite $g$ and they
are then protected by the glide symmetry.
Furthermore, we find that in the one-dimensional (1D) cuts at any given $k_2$ the system has an inversion topological
number $\mathbb{Z}^\ge$ \cite{Ber14}, with the inversion symmetry expressed by 
${\cal R}_{k_{2}}^{t}$. $\mathbb{Z}^\ge$ is defined
as the difference of the number of occupied states
with a chosen inversion eigenvalue at the two inversion invariant points, $k_{1}=0$
and $k_{1}=\pi$.
Taking the spectra in the $g_{+}$ glide sector at $k_{1}=\{0,\pi\}$, one can immediately deduce that
$\mathbb{Z}^\ge$ changes sign at the position of the DPS, i.e. $k^{0}$ and $2\pi-k^{0}$.
The inversion symmetry protection of the DPs explains the presence of a third edge state close to the zone boundary 
in Fig. \ref{fig:z2_sm}(c). In the absence of glide symmetry, $\mathbb{Z}^\ge$ is meaningful only
in the two high-symmetry cuts of the BZ, i.e. $k_{2}=\{0,\pi\}$, and it is non-trivial at $k_{2}=0$ but trivial 
at $k_{2}=\pi$, so there must be gap
closings between these two 1D cuts (Fig. \ref{fig:patt_PDs}(f)).
Finally, the $z2$ state exhibits also a symmetry protection arising from the combination of ${\cal I}$ and $\cal{T}$. Their product
yields a conjugation operator, ${\cal K}_{\vec{k}}\equiv{\cal I}_{\vec{k}}{\cal T}$,
since ${\cal K}$ transforms ${\cal H}_{\vec{k}}$ into ${\cal H}^T_{\vec{k}}$. 
If ${\cal K}_{\vec{k}}{\cal K}_{\vec{k}}^{\star}\equiv1$
(which is satisfied for any $L_{z}$) then ${\cal H}$ can be made purely real 
in the eigenbasis of ${\cal K}_{\vec{k}}$ (as demonstrated explicitly in the Appendix \ref{sec:re_ham}). 
Indeed, close to the DPs, the low energy
Hamiltonian has a form $H_{\delta \vec{k}}=\delta k_{1}A+\delta k_{2}B$
with $A$ and $B$ being $2\times2$ real matrices and $\delta\vec{k}$
the deviation with respect to the DP. The charge can be then calculated
as a $\mathbb{Z}_{2}^{(1)}$ mod-$2$ winding number (see Appendix \ref{sec:topo_inv}), and it takes values $\pm1$ at the two DPs. Though the time-inversion operator ${\cal K}$ is explicitly dependent on the momentum, in the low-energy description the result is consistent with the general expectation \cite{Zha2016} of having a non-trivial $\mathbb{Z}_2$ topological
invariant at the DPs by combination of time and inversion symmetry.   
Hence, the resulting electronic spectra for the $z2$ phase 
in Figs. \ref{fig:z2_sm} (b)-(e) provide evidence of a peculiar
{\it topological resilience} of DPs and edge modes, being robust to different types of 
symmetry breaking configurations.

Concerning the lowering symmetry perturbations, we point out that breaking of glide 
in such a way that inversion
is not preserved opens a gap and the system becomes topologically
trivial, consistently with the expectations from the classification table for the AI class in the 
presence of a mirror reflection \cite{Sch14}. Otherwise, breaking of the reflection or time only removes
the $\mathbb{Z}_{2}$ protection but leaves the glide symmetry so that the
DPs are still preserved. In Fig. \ref{fig:z2_sm}(e) a representative case
of time-reversal violation is also considered. Due to a termination dependent orbital polarization, non-degenerate chiral 
states emerge at the edge and, remarkably, the system allows for non-vanishing charge and orbital currents at the boundary. 
\begin{figure}[t]
\includegraphics[clip,width=1\columnwidth]{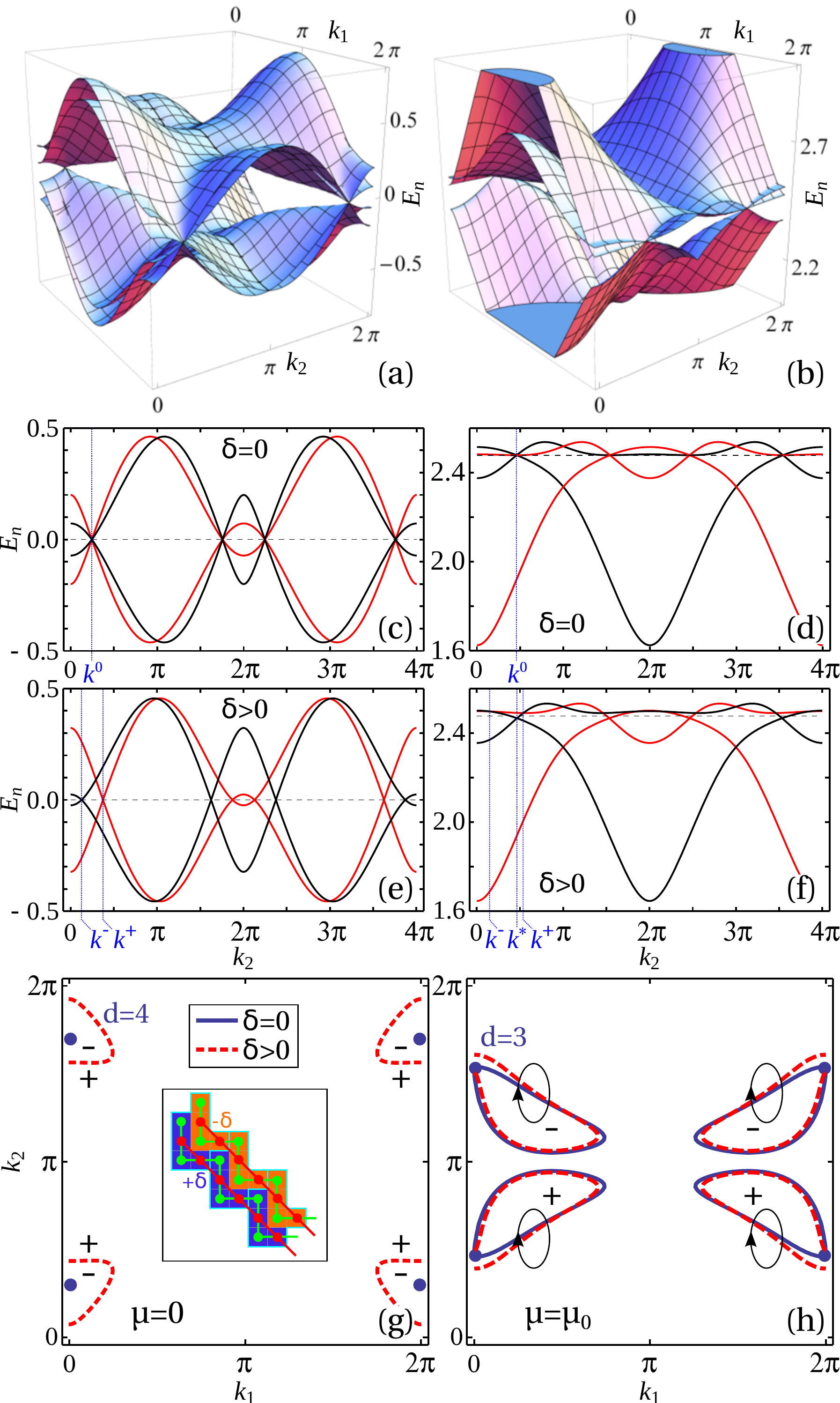}\protect\caption{
Multiple FPs in $z3$ antiferromagnet. (a) bands with chiral four-fold
DPs at half filling ($\mu=0$), (b) bands with three-fold
Fermi points at higher filling ($\mu=\mu_{0}$). (c), (d) the same
bands in the glide plane $k=0$ with color indicating the glide eigenvalue
$+1$ (red) and $-1$ (black). Dashed line is the Fermi level $\mu$.
(e), (f) as in (c) and (d) plus a symmetry conserving long-range hopping $\delta$. 
(g), (h) FSs containing
multiple FPs at $\delta=0$ (blue) and $\delta>0$ (red
dashed). Multiple band crossing points are marked with dots and their degeneracy is
indicated. The inset of (g) shows a schematic view of the hopping
$\delta$. The signs in (g) indicate the band parity,
$+1$ inside the Fermi pockets and $-1$ outside. The signs in (h) are
those due the AI winding numbers of the Fermi lines. The integration contours
are indicated by ellipses.\label{fig:z3_mult}}
\end{figure}

Let us consider the $z3$ AFM for two representative electron densities; at half-filling and away from it, at 
$\mu=\mu_0$.
Firstly, we would like to focus on the consequence of the glide symmetry on the electronic spectra by investigating the origin of the observed multiple band touchings in the glide plane,
as depicted in the Figs. \ref{fig:z3_mult}(a),(b).
Indeed, the gapless phases in the 
GPs for $z3$ AFM can exhibit quadruple and triple band touchings
at half-filling and away from it, respectively (Figs. \ref{fig:z3_mult}(a),(b)). 
As we will discuss more extensively later on, a prerequisite for having multiple Fermi points is to get DPs that are topologically protected in the glide plane.
Furthermore, we find that the DPs can get glued and that the multiplicity of the FPs 
is not accidental but it also arises as a consequence of the glide symmetry. 
Their origin the GPs is due to a {\it hidden} non-unitary relation
that leaves invariant the determinant of the glide-block Hamiltonian.
One can demonstrate (see Appendix \ref{sec:shift_glide}) that for the Hamiltonian projected in the GP 
the two blocks are related by a $2\pi$ shift, i.e.,
$H_{0,k_{2}}^{+}=H_{0,k_{2}+2\pi}^{-}$. 
However, while the spectrum of $H_{0,k_{2}}^{+}$
is $4\pi$-periodic, due to the non-primitive half-lattice translation of the glide symmetry, 
its determinant has a $2\pi$-periodicity, 
$\rm{det}(H_{0,k_{2}}^{+}-\mu)\equiv \rm{det}(H_{0,k_{2}+2\pi}^{+}-\mu)$. It is important noticing that such 
property of the determinant does not depend on the chiral symmetry of the spectrum, because it manifests both
at $\mu=0$ and $\mu=\mu_{0}=\sqrt{2+J_{H}^{2}+\lambda^{2}}$, respectively. 
We argue that such feature is
remnant of the original periodicity of the full Hamiltonian and it
implies that a $\mu$ eigenvalue of $H_{0,k_{2}}^{+}$
must be $2\pi$ periodic. Then, ($i$) the $2\pi$-shift
relation of the glide symmetric blocks and ($ii$) the $2\pi$-periodic determinant 
combine to glue together DPs from different glide-blocks at two specific energy levels, 
$\mu=0,\mu_0$. We point out that while such property has similarity with the 
{\it congruence}, it is not exactly equivalent. Indeed, only a hermitian congruence is found between the blocks
$H^+_{0,k_2}$ and $H^+_{0,k_2+2\pi}$, thus meaning that in a given basis
these blocks differ only by a scaling of the rows and columns entries \cite{Rem}. 
The anomalous periodicity of the determinant can be constructed explicitly, as demonstrated in the Appendix \ref{sec:det_eq}, 
through a non-unitary chiral-like operator $\Sigma_{k_{2}}\equiv h_{k_{2}}^{-1}\bar{H}{}_{k_{2}}$
where $\bar{H}{}_{k_{2}}$ is the $2\pi$ periodic part of $H_{0,k_{2}}^{+}$ and
$h_{k_{2}}$ is the part with $4\pi$ period ($h_{k_{2}+2\pi}\equiv-h_{k_{2}}$). 
The proof relies on the fact that eigenvalues of $\Sigma_{k_{2}}$ are symmetric around
zero and on the Silvester property of determinants, i.e., $\rm{det}(1+AB)\equiv \rm{det}(1+BA)$.
The emergence of a non-unitary relation that leaves the determinant invariant  
represents a novel mechanism for the search and generation of multiple band touching points in the presence 
of NS symmetries.
  
When breaking the conditions for the determinant invariance in the glide sectors, the DPs split into simple DPs.
To show this behavior we add a symmetry conserving second and third nearest-neighbors hopping ($\delta$) 
in order to remove only the DPs degeneracy (Figs. \ref{fig:z3_mult}(e),(f)).
Such removal is concomitant with a transition from semimetal to metal whose Fermi surface (FS) is topological protected and has a structure which is strongly
tied to the presence of topological non-trivial DPs in the glide plane. 
At $\mu=0$ and finite $\delta$ the FS has a topological character which arises from the glide symmetry and the induced non-symmorphic nature of the
chiral transformation. Indeed, by combining the chiral and the conjugation symmetries
we find a $k$-dpendent anticonjugation operator ${\cal A}_{\vec{k}}={\cal S}_{k_{1}}{\cal K}_{\vec{k}}$,
which act to give a minus complex conjugate of ${\cal H}_{\vec{k}}$ (see Appendix \ref{sec:imag_ham}).
One can show that ${\cal A}_{\vec{k}}{\cal A}_{\vec{k}}^{\star}\equiv1$
for any odd $L_{z}$ and can be also rearranged as a product of the $k$-dependent inversion and particle-hole symmetry operators. We point out that due to its $k$-dependence, such symmetry  
is not explicitly included in the topological classification of a 1D FS by combination of particle-hole and inversion \cite{Zha2016}.
Taking into account the structure of the ${\cal A}$
operator, one can construct a $\mathbb{Z}_2$ invariant by putting the Hamiltonian in the antisymmetric form using the
eigenbasis of ${\cal A}_{\vec{k}}$ and calculating its Pfaffian (see Appendix \ref{sec:imag_ham}).
The main result is that at the FS the Pfaffian changes sign and band inversion occurs (Fig. \ref{fig:z3_mult} (g)) thus the FS can be topologically protected.
Apart from the whole FS, one can notice that in the GP, at $k_1=0$, the ${\cal A}$ operator is not $k$-dependent because the chirality depends only from $k_1$, and, thus, it also provides a protection for the DPs in both glide sectors. Then, if the metallic phase can be converted into a semimetal state by suitably tuning the microscopic parameters, the transition can only occur in the glide plane and the semimetal phase will exhibit multiple degenerate DPs, as it happens for instance at $\delta=0$. Such outcome can be generally expected for glide and chiral symmetric metallic phases in AI class which exhibit a topological protection in the full Brillouin zone and in one of the glide planes.   
Breaking of the glide symmetry opens a gap at the Fermi level everywhere in the Brillouin zone except that in the glide plane where DPs can be still protected by the conventional particle-hole and inversion combination \cite{Zha2016} as it is observed for the investigated electronic phases at half-filling. 
Hence, the emerging metallic phase is topologically protected and is marked by two Fermi pockets that 
are {\it glued} to the glide plane through the symmetry protected DPs.
Away from half-filling, e.g. at $\mu=\mu_0$, the metallic phase is protected by the time symmetry as it exhibits a non trivial AI winding number $\mathbb{Z}$(see Appendix for the details of the calculation).  In general we cannot have a semimetal phase and the triple band crossing coexists with a Fermi pocket (Fig. \ref{fig:z3_mult} (e)). The removal of the degeneracy by $\delta$ leads to an AI topological metal phase.

\section{Conclusions \label{sec:konk}} 
 
The realization of topological zig-zag AFMs and edge states has been so far focusing 
on insulating configurations \cite{Li2016,Brey2005,Du2015}. Our analysis demonstrates that the breakdown of the fully gapped phases in a multi-orbital zig-zag AFMs, due to the NS symmetry, can lead to topological gapless phases whose nature depends on the characteristic zig-zag length as well as on the intricate entanglement of non-symmorphic and internal or other spatial symmetries. 
The results demonstrate that, for a generic class of two-dimensional zig-zag AFMs with collinear magnetic order in the AI symmetry class, gapless phases are prone to exhibit a topological behavior.
We find that the invariance of the determinants in the glide sectors is a novel mechanism, uniquely 
arising in NS systems, to generate multiple DPs and topological non-trivial gapless phases.
Due to the observed $\mathbb{Z}_2$ protection of the Fermi points and the strong nesting of the topological metallic phase it is plausible to expect that the metallic phase can exhibit an anomalous magnetotransport response as well as a tendency to other electronic instabilities.

Concerning the materials perspective, there are many compounds exhibiting zig-zag magnetic patterns that involve $t_{2g}$ orbitals close to the Fermi level especially when considering transition metal oxides.
In this framework, our results may find interesting application both in Mn doped ruthenates and dichalcogenides. 

%
%
{\bf Acknowledgements} W.B. acknowledges support by the European Horizon 2020 research and innovation programme under 
the Marie-Sklodowska-Curie grant agreement No. 655515
and by National Science Center (NCN) under Project No. 2012/04/A/ST3/00331.

\appendix

\section{Stucture of the Hamiltonian for $L_{z}=2$ and $L_{z}=3$ zigzag patterns \label{sec:struc}}

The general form of the Hamiltonian in Eq. 1 (see main text) for a zigzag spin pattern
can be obtained by analyzing the possible hopping processes for a given unit
cell, as shown in Fig. 1 (see the main text). Its matrix representation for
a given quasimomentum $\vec{k}$ and fixed spin polarization of the
itinerant electrons, that is a good quantum number, can be conveniently
written by a block matrix in the form 
\begin{equation}
{\cal H}_{\vec{k}}=\begin{pmatrix}
{\bf H}_{\downarrow\downarrow}^{bb} & {\bf H}_{\downarrow\uparrow}^{bb} & {\bf H}_{\downarrow\downarrow}^{ba} & {\bf H}_{\downarrow\uparrow}^{ba} \\ {\bf H}_{\uparrow\downarrow}^{bb} & {\bf H}_{\uparrow\uparrow}^{bb} & {\bf H}_{\uparrow\downarrow}^{ba} & {\bf H}_{\uparrow\uparrow}^{ba} \\ {\bf H}_{\downarrow\downarrow}^{ab} & {\bf H}_{\downarrow\uparrow}^{ab} & {\bf H}_{\downarrow\downarrow}^{aa} & {\bf H}_{\downarrow\uparrow}^{aa} \\ {\bf H}_{\uparrow\downarrow}^{ab} & {\bf H}_{\uparrow\uparrow}^{ab} & {\bf H}_{\uparrow\downarrow}^{aa} & {\bf H}_{\uparrow\uparrow}^{aa} \end{pmatrix} .
\label{eq:block_ham}
\end{equation}

Here, the blocks ${\bf H}_{\sigma\sigma'}^{\alpha\beta}$($\sigma,\sigma'=\uparrow,\downarrow$
and $\alpha,\alpha'=a,b$) are of size $N_{\downarrow}=N_{\uparrow}=2L_{z}-2$,
associated to the spin up and down domains within the unit cell (see
Fig. 1). The indices $(\sigma\alpha)$ and $(\sigma^{'}\beta)$
mean that the block describes hopping from the spin $\sigma$ to spin
$\sigma'$ domains and orbitals $\alpha$ and $\beta$, respectively. For
$z2$ magnetic pattern the blocks for the electrons with spin up
are given by the equations
\begin{eqnarray}
{\bf H}_{\downarrow\downarrow}^{bb} & = & 
\begin{pmatrix} J_{H} & -e^{-ik_{2}} t_{\hat{a}}^{bb}+t_{\hat{b}}^{bb}\\
-e^{ik_{2}} t_{\hat{a}}^{bb}+t_{\hat{b}}^{bb} & J_{H}
\end{pmatrix},\nonumber\\
{\bf H}_{\uparrow\uparrow}^{bb} & = & 
\begin{pmatrix} -J_{H} & -e^{-ik_{2}} t_{\hat{a}}^{bb}+t_{\hat{b}}^{bb}\\
-e^{ik_{2}} t_{\hat{a}}^{bb}+t_{\hat{b}}^{bb} & -J_{H}
\end{pmatrix},\nonumber\\
{\bf H}_{\downarrow\uparrow}^{bb} & = & 
\begin{pmatrix} 0 & e^{-ik_{2}} t_{\hat{b}}^{bb} -t_{\hat{a}}^{bb}\\
-e^{-ik_{1}} t_{\hat{a}}^{bb}+e^{i(k_{2}-k_{1})} t_{\hat{b}}^{bb} & 0
\end{pmatrix},\nonumber\\
{\bf H}_{\uparrow\downarrow}^{bb} & = & \left({\bf H}_{\downarrow\uparrow}^{bb}\right)^{\dagger},
\end{eqnarray}
for the $b$-orbital sector and
\begin{eqnarray}
{\bf H}_{\downarrow\downarrow}^{aa} & = & 
\begin{pmatrix}J_{H} & e^{-ik_{2}}t_{\hat{a}}^{aa}-t_{\hat{b}}^{aa}\\
e^{ik_{2}}t_{\hat{a}}^{aa}-t_{\hat{b}}^{aa} & J_{H}
\end{pmatrix},\nonumber\\
{\bf H}_{\uparrow\uparrow}^{aa} & = & 
\begin{pmatrix}-J_{H} & e^{-ik_{2}}t_{\hat{a}}^{aa}-t_{\hat{b}}^{aa}\\
e^{ik_{2}}t_{\hat{a}}^{aa}-t_{\hat{b}}^{aa} & -J_{H}
\end{pmatrix},\nonumber\\
{\bf H}_{\downarrow\uparrow}^{aa} & = & 
\begin{pmatrix}0 & -e^{-ik_{2}}t_{\hat{b}}^{aa}+t_{\hat{a}}^{aa}\\
e^{-ik_{1}}t_{\hat{a}}^{aa}-e^{i(k_{2}-k_{1})}t_{\hat{b}}^{aa} & 0
\end{pmatrix},\nonumber\\
{\bf H}_{\uparrow\downarrow}^{aa} & = & \left({\bf H}_{\downarrow\uparrow}^{aa}\right)^{\dagger},
\end{eqnarray}
 for the $a$-orbital sector. For the interorbital sector we have,
\begin{eqnarray}
{\bf H}_{\downarrow\downarrow}^{ba} & = & \begin{pmatrix}i\lambda & e^{-ik_{2}}t_{\hat{a}}^{ab}+t_{\hat{b}}^{ab}\\
-e^{ik_{2}}t_{\hat{a}}^{ab}-t_{\hat{b}}^{ab} & i\lambda
\end{pmatrix},\nonumber\\
{\bf H}_{\uparrow\uparrow}^{ba} & = & \begin{pmatrix}-i\lambda & e^{-ik_{2}}t_{\hat{a}}^{ab}+t_{\hat{b}}^{ab}\\
-e^{ik_{2}}t_{\hat{a}}^{ab}-t_{\hat{b}}^{ab} & -i\lambda
\end{pmatrix},\nonumber\\
{\bf H}_{\downarrow\uparrow}^{ba} & = & \begin{pmatrix}0 & e^{-ik_{2}}t_{\hat{b}}^{ab}+t_{\hat{a}}^{ab}\\
-e^{-ik_{1}}t_{\hat{a}}^{ab}-e^{i(k_{2}-k_{1})}t_{\hat{b}}^{ab} & 0
\end{pmatrix},\nonumber\\
{\bf H}_{\uparrow\downarrow}^{ba} & = & \begin{pmatrix}0 & e^{ik_{1}}t_{\hat{a}}^{ab}+e^{i(k_{1}-k_{2})}t_{\hat{b}}^{ab}\\
-e^{ik_{2}}t_{\hat{b}}^{ab}-t_{\hat{a}}^{ab} & 0
\end{pmatrix},
\end{eqnarray}
and the rest of blocks can be recovered from these as 
\begin{equation}
{\bf H}_{\sigma\sigma'}^{ab}=\left({\bf H}_{\sigma'\sigma}^{ba}\right)^{\dagger}.\label{eq:Hdag}
\end{equation}
We note that the spin sectors for the itinerant electrons are completely
equivalent. It is immediate to verify that the Hamiltonian blocks for the
opposite spin-sectors can be obtained by changing the sign of $J_{H}$ and $\lambda$. For
zig-zag $z3$ the size of the spin domain is $N_{\downarrow}=4$ so that 
the blocks are twice larger. Then, we have
\begin{eqnarray}
{\bf H}_{\downarrow\downarrow}^{bb} & = & \begin{pmatrix}J_{H} & t_{\hat{b}}^{bb} & 0 & -e^{-ik_{2}}t_{\hat{a}}^{bb}\\
t_{\hat{b}}^{bb} & J_{H} & t_{\hat{b}}^{bb} & 0\\
0 & t_{\hat{b}}^{bb} & J_{H} & -t_{\hat{a}}^{bb}\\
-e^{ik_{2}}t_{\hat{a}}^{bb} & 0 & -t_{\hat{a}}^{bb} & J_{H}
\end{pmatrix},\nonumber\\
{\bf H}_{\uparrow\uparrow}^{bb} & = & \begin{pmatrix}-J_{H} & t_{\hat{b}}^{bb} & 0 & -e^{-ik_{2}}t_{\hat{a}}^{bb}\\
t_{\hat{b}}^{bb} & -J_{H} & t_{\hat{b}}^{bb} & 0\\
0 & t_{\hat{b}}^{bb} & -J_{H} & -t_{\hat{a}}^{bb}\\
-e^{ik_{2}}t_{\hat{a}}^{bb} & 0 & -t_{\hat{a}}^{bb} & -J_{H}
\end{pmatrix},\nonumber\\
{\bf H}_{\downarrow\uparrow}^{bb} & = & N_{\downarrow}\begin{pmatrix}0 & -t_{\hat{a}}^{bb} & 0 & e^{-ik_{2}}t_{\hat{b}}^{bb}\\
-e^{-ik_{1}}t_{\hat{a}}^{bb} & 0 & -t_{\hat{a}}^{bb} & 0\\
0 & -e^{-ik_{1}}t_{\hat{a}}^{bb} & 0 & e^{-ik_{1}}t_{\hat{b}}^{bb}\\
e^{i(k_{2}-k_{1})}t_{\hat{b}}^{bb} & 0 & t_{\hat{b}}^{bb} & 0
\end{pmatrix},\nonumber\\
{\bf H}_{\uparrow\downarrow}^{bb} & = & \left({\bf H}_{\downarrow\uparrow}^{bb}\right)^{\dagger},
\end{eqnarray}
for the $b$-orbital sector, and 
\begin{eqnarray}
{\bf H}_{\downarrow\downarrow}^{aa} & = & \begin{pmatrix}J_{H} & -t_{\hat{b}}^{aa} & 0 & e^{-ik_{2}}t_{\hat{a}}^{aa}\\
-t_{\hat{b}}^{aa} & J_{H} & -t_{\hat{b}}^{aa} & 0\\
0 & -t_{\hat{b}}^{aa} & J_{H} & t_{\hat{a}}^{aa}\\
e^{ik_{2}}t_{\hat{a}}^{aa} & 0 & t_{\hat{a}}^{aa} & J_{H}
\end{pmatrix},\nonumber\\
{\bf H}_{\uparrow\uparrow}^{aa} & = & \begin{pmatrix}-J_{H} & -t_{\hat{b}}^{aa} & 0 & e^{-ik_{2}}t_{\hat{a}}^{aa}\\
-t_{\hat{b}}^{aa} & -J_{H} & -t_{\hat{b}}^{aa} & 0\\
0 & -t_{\hat{b}}^{aa} & -J_{H} & t_{\hat{a}}^{aa}\\
e^{ik_{2}}t_{\hat{a}}^{aa} & 0 & t_{\hat{a}}^{aa} & -J_{H}
\end{pmatrix},\nonumber\\
{\bf H}_{\downarrow\uparrow}^{aa} & = & \begin{pmatrix}0 & t_{\hat{a}}^{aa} & 0 & -e^{-ik_{2}}t_{\hat{b}}^{aa}\\
e^{-ik_{1}}t_{\hat{a}}^{aa} & 0 & t_{\hat{a}}^{aa} & 0\\
0 & e^{-ik_{1}}t_{\hat{a}}^{aa} & 0 & -e^{-ik_{1}}t_{\hat{b}}^{aa}\\
-e^{i(k_{2}-k_{1})}t_{\hat{b}}^{aa} & 0 & -t_{\hat{b}}^{aa} & 0
\end{pmatrix},\nonumber\\
{\bf H}_{\uparrow\downarrow}^{aa} & = & \left({\bf H}_{\downarrow\uparrow}^{aa}\right)^{\dagger},
\end{eqnarray}
 for the $a$-orbital sector. For the interorbital sector we have
\begin{eqnarray}
{\bf H}_{\downarrow\downarrow}^{ba} & = & \begin{pmatrix}i\lambda & t_{\hat{b}}^{ab} & 0 & e^{-ik_{2}}t_{\hat{a}}^{ab}\\
-t_{\hat{b}}^{ab} & i\lambda & -t_{\hat{b}}^{ab} & 0\\
0 & t_{\hat{b}}^{ab} & i\lambda & t_{\hat{a}}^{ab}\\
-e^{ik_{2}}t_{\hat{a}}^{ab} & 0 & -t_{\hat{a}}^{ab} & i\lambda
\end{pmatrix},\\
{\bf H}_{\uparrow\uparrow}^{ba} & = & \begin{pmatrix}-i\lambda & t_{\hat{b}}^{ab} & 0 & e^{-ik_{2}}t_{\hat{a}}^{ab}\\
-t_{\hat{b}}^{ab} & -i\lambda & -t_{\hat{b}}^{ab} & 0\\
0 & t_{\hat{b}}^{ab} & -i\lambda & t_{\hat{a}}^{ab}\\
-e^{ik_{2}}t_{\hat{a}}^{ab} & 0 & -t_{\hat{a}}^{ab} & -i\lambda
\end{pmatrix},\nonumber\\
{\bf H}_{\downarrow\uparrow}^{ba} & = & \begin{pmatrix}0 & t_{\hat{a}}^{ab} & 0 & e^{-ik_{2}}t_{\hat{b}}^{ab}\\
-e^{-ik_{1}}t_{\hat{a}}^{ab} & 0 & -t_{\hat{a}}^{ab} & 0\\
0 & e^{-ik_{1}}t_{\hat{a}}^{ab} & 0 & e^{-ik_{1}}t_{\hat{b}}^{ab}\\
-e^{i(k_{2}-k_{1})}t_{\hat{b}}^{ab} & 0 & -t_{\hat{b}}^{ab} & 0
\end{pmatrix},\nonumber\\
{\bf H}_{\uparrow\downarrow}^{ba} & = & \begin{pmatrix}0 & e^{ik_{1}}t_{\hat{a}}^{ab} & 0 & e^{i(k_{1}-k_{2})}t_{\hat{b}}^{ab}\\
-t_{\hat{a}}^{ab} & 0 & -e^{ik_{1}}t_{\hat{a}}^{ab} & 0\\
0 & t_{\hat{a}}^{ab} & 0 & t_{\hat{b}}^{ab}\\
-e^{ik_{2}}t_{\hat{b}}^{ab} & 0 & -e^{ik_{1}}t_{\hat{b}}^{ab} & 0
\nonumber
\end{pmatrix},
\end{eqnarray}
and the remaining blocks are as those in Eq. (\ref{eq:Hdag})
\begin{equation}
{\bf H}_{\sigma\sigma'}^{ab}=\left({\bf H}_{\sigma'\sigma}^{ba}\right)^{\dagger}.
\end{equation}
On adding the second and third neighbor hopping $\delta$ the blocks
${\bf H}{}_{\sigma\sigma}^{\alpha\alpha}$ of the zig-zag $3$ are
modified in a following way, ${\bf H}{}_{\sigma\sigma}^{\alpha\alpha}\to{\bf H}{}_{\sigma\sigma}^{\alpha\alpha}+{\bf h}(\delta)_{\sigma\sigma}^{\alpha\alpha}$
with, 
\begin{equation}
{\bf h}_{\sigma\sigma}^{\alpha\alpha}=\delta\sigma\begin{pmatrix}0 & 0 & 1\!+\!e^{-ik_{2}} & 0\\
0 & 0 & 0 & 1\!+\!e^{-ik_{2}}\\
1\!+\!e^{ik_{2}} & 0 & 0 & 0\\
0 & 1\!+\!e^{ik_{2}} & 0 & 0
\end{pmatrix}.
\end{equation}
From now on we will assume hopping amplitudes for undistorted (cubic) system,
 i.e., $t_{\hat{a},bb}=t_{\hat{b},aa}=-t$ and $t_{\hat{\gamma},ab}=0$. 

\section{Symmetry properties}

In this section we will present the explicit expressions of the symmetry operators and all the consequences on their structure as related to the presence of the non-symmorphic glide symmetry.

\subsection{Non-spatial symmetries\label{sec:non-spa_sym}}

\begin{figure}[t]
\includegraphics[clip,width=1\columnwidth]{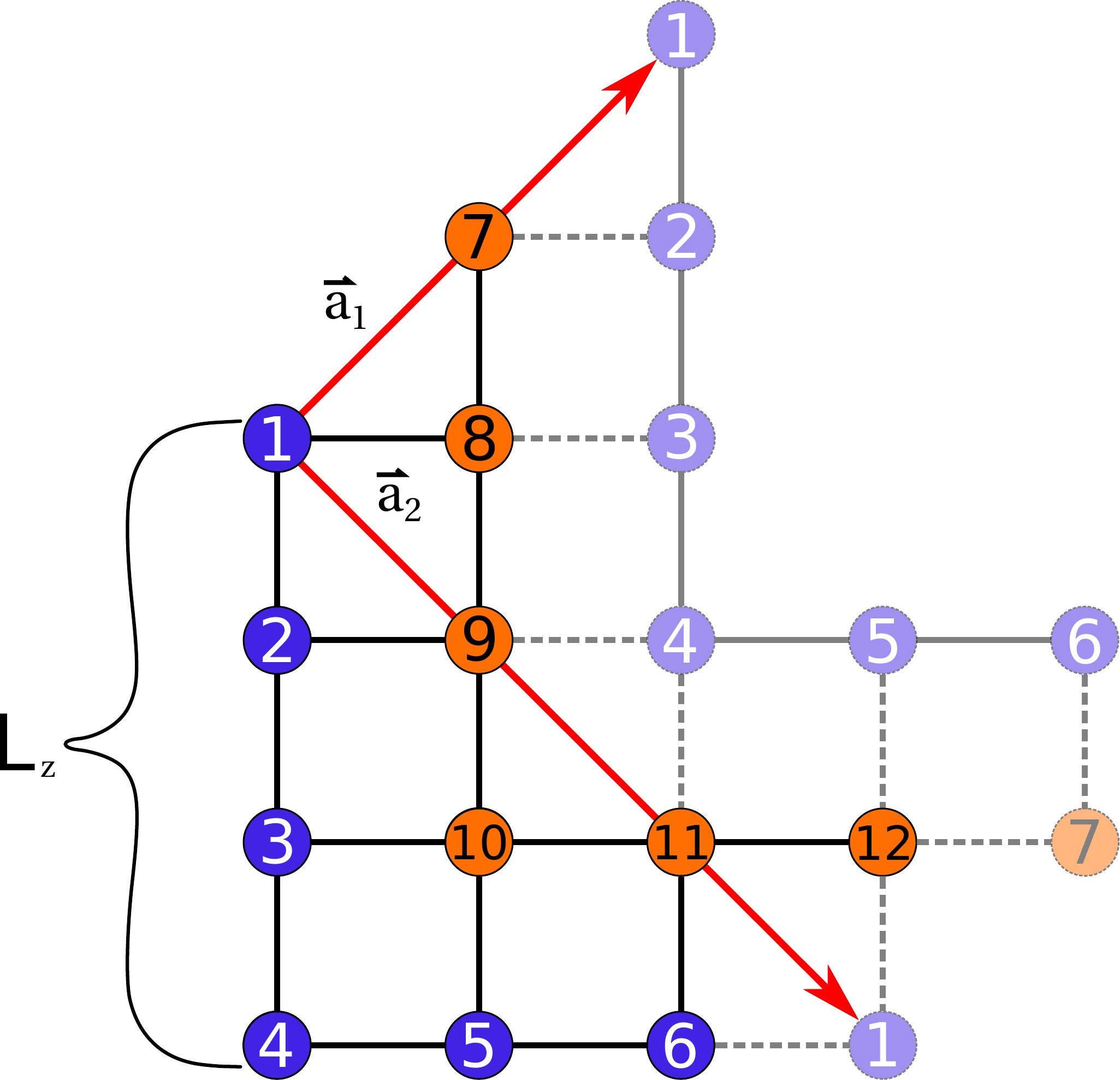}
\protect\caption{Schematic view of the unit cell for zig-zag
length $L_{z}=4$ and labeling of the intra-cell sites. Orange and blue circles indicate the sites with spin
up and down, respectively. Solid lines connect sites belonging to one unit cell
whereas the dashes line connect sites of neighboring unit cells. The
primitive lattice translation vectors $\vec{a}_{1,2}$ are shown explicitly.\label{fig:cell}}
\end{figure}
The spin orientation of the itinerant electrons
is a good quantum number for the problem upon examination.
In order to find the time reversal operator, 
one needs to get a unitary matrix $T$ that satisfies
the relation
\begin{equation}
{\cal T}^{\dagger}{\cal H}_{\vec{k}}{\cal T}={\cal H}_{-\vec{k}}^{T}.
\end{equation}
Note that the transposition on the {\it rhs} is equivalent to complex conjugation,
so that the whole transformation is antiunitary. Looking at
a given block of ${\cal H}_{\vec{k}}$ for $z2$ and
$z3$ zig-zag patterns, it is direct to verify that the complex conjugation will only change
all the $\vec{k}$ to $-\vec{k}$ and change $\lambda$ to $-\lambda$.
If we consider the two orbital flavors as the two independent effective {\it layers} 
of the square lattice then in the absence of the inter-orbital
hopping, i.e, when no distortion is present, the change of sign in
$\lambda$ can be absorbed by a gauge transformation of the form
\begin{eqnarray}
d_{i,a,\sigma}^{(\dagger)} & = & -\widetilde{d}_{i,a,\sigma}^{(\dagger)}\nonumber \\
d_{i,b,\sigma}^{(\dagger)} & = & \tilde{d}_{i,b,\sigma}^{(\dagger)}\label{eq:gaug1}
\end{eqnarray}
where $\tilde{d}^{(\dagger)}$ are the new fermion operators. Since
all the hopping processes are orbital-conserving, i.e., there is no effective inter-orbital
mixing, the only effect of this gauge transformation is to change
the sign of $\lambda$. Thus, the form of the operator
${\cal T}$ is
\begin{equation}
{\cal T}=\begin{pmatrix}{\bf 1}_{\downarrow}^{b} & {\bf 0} & {\bf 0} & {\bf 0}\\
{\bf 0} & {\bf 1}_{\uparrow}^{b} & {\bf 0} & {\bf 0}\\
{\bf 0} & {\bf 0} & -{\bf 1}_{\downarrow}^{a} & {\bf 0}\\
{\bf 0} & {\bf 0} & {\bf 0} & -{\bf 1}_{\uparrow}^{a}
\end{pmatrix},
\end{equation}
where ${\bf 1}$ denotes a unit matrix of the size $N_{\downarrow}$
and the indices are left to indicate spin/orbital sector. Since the
matrix is purely real we also observe that $T^{2}=1$. 

Another important
non-spatial symmetry is that one associated to the sublattice or chiral-like symmetry. 
Nevertheless, due to the intrinsic antiferromagnetic structure of the zig-zag pattern, one can find 
a unitary operator ${\cal S}_{k_{1}}$ that anticommutes
with Hamiltonian but is explicitly momentum dependent as 
\begin{equation}
{\cal S}_{k_{1}}^{\dagger}{\cal H}_{\vec{k}}{\cal S}_{k_{1}}=-{\cal H}_{\vec{k}} \,.
\end{equation}
In our case such symmetry occurs only at half-filling and arises from
the two sublattice structure of the two magnetic domains
within a unit cell as marked with different colors in Fig. \ref{fig:cell}.
As one can deduce from the figure, in order to move from one domain to another one, 
a translation $\vec{a}_{1}$ is needed. For this reason, the operator $S_{k_{1}}$
depends on the quasimomentum $k_{1}$ and has non-zero blocks that
connect opposite spin domains within the same orbital sectors, i.e.,
\begin{equation}
{\cal S}_{k_{1}}=\begin{pmatrix}{\bf 0} & {\bf S}_{k_{1}} & {\bf 0} & {\bf 0}\\
{\bf S}_{k_{1}}^{\dagger} & {\bf 0} & {\bf 0} & {\bf 0}\\
{\bf 0} & {\bf 0} & {\bf 0} & {\bf S}_{k_{1}}\\
{\bf 0} & {\bf 0} & {\bf S}_{k_{1}}^{\dagger} & {\bf 0}
\end{pmatrix},
\end{equation}
with 
\begin{equation}
{\bf S}_{k_{1}}=e^{-i\frac{k_{1}}{2}}\begin{pmatrix}-1 & 0 & 0 & \cdots & 0\\
0 & 1 & 0 & \cdots & 0\\
0 & 0 & -1 & \cdots & 0\\
\vdots & \vdots & \vdots & \ddots & 0\\
0 & 0 & 0 & 0 & 1
\end{pmatrix}.
\end{equation}
As we see, apart from the change of the spin-domain there is a phase
factor of $e^{-i\frac{k_{1}}{2}}$ appearing with alternating sign
$(-1)^{i}$ as we go along the zig-zag segment (from site $i=1$ to $i=8$
in Fig. \ref{fig:cell}). The alternation is necessary to change the
signs of the allowed hoppings that are along the bonds of the square lattice.
Here, the chiral symmetry is present only because the zig-zag
sublattice structure of the magnetic domains is compatible with the natural
two-sublattice structure of the square lattice. Finally, the last
non-spatial symmetry is the particle-hole (PHS) of charge conjugation
symmetry. As for the chirality, the particle-hole operator is also momentum dependent and 
satisfies the relation
\begin{equation}
{\cal C}_{k_{1}}^{\dagger}{\cal H}_{\vec{k}}{\cal C}_{k_{1}}=-{\cal H}_{-\vec{k}}^{T}.\label{eq:phs}
\end{equation}
Such condition implies that if both TRS and sublattice symmetry exist
then the charge conjugation is their product, ${\cal C}_{k_{1}}={\cal S}_{k_{1}}\cdot{\cal T}$.
It is direct to check that despite the $k$ - dependence in ${\cal S}_{k_{1}}$
we have ${\cal C}^{2}=1$ (in case of the $k$-dependent antiunitary
operator this means ${\cal C}_{k_{1}}{\cal C}_{-k_{1}}^{\star}\equiv1$,
where star is complex conjugation). 


\subsection{Spatial symmetries in a cubic system\label{sec:spa_sym}}

The reflection plane for a zig-zag pattern with size $L_{z}=2,3$
is shown in Fig. \ref{fig:patt_PDs}(a)-(b) and can be easily constructed for any $L_{z}$.
Its direction is diagonal with respect to the cubic axes so it interchanges
the hoppings along the $a$ and $b$ symmetry directions. Such transformation, then, 
requires an interchange
of the orbitals $a$ and $b$ to preserve the connectivity of the system
before performing the reflection. Finally, such interchange effectively modifies
the sign of $\lambda$ so one can again introduce a gauge transformation
as given by Eq. (\ref{eq:gaug1}). Since the elementary cell of the zig-zag
system shown in Fig. \ref{fig:patt_PDs}(a)-(b) is not left invariant by the
reflection (one can fix it by an alternative definition of the elementary
cell or by taking the square unit cell) the reflection operator depends
on quasimomentum $k_{2}$ and has the form
\begin{equation}
{\cal R}_{k_{2}}=\begin{pmatrix}{\bf 0} & {\bf 0} & {\bf R}_{k_{2}} & {\bf 0}\\
{\bf 0} & {\bf 0} & {\bf 0} & {\bf {\bf R}}_{k_{2}}\\
-{\bf {\bf R}}_{k_{2}} & {\bf 0} & {\bf 0} & {\bf 0}\\
{\bf 0} & -{\bf {\bf R}}_{k_{2}} & {\bf 0} & {\bf 0}
\end{pmatrix},
\end{equation}
with $N_{\downarrow}\times N_{\downarrow}$ blocks given by
\begin{equation}
{\bf R}_{k_{2}}=\begin{pmatrix}i & 0 & 0 & \cdots & 0 & 0\\
0 & 0 & 0 & \cdots & 0 & ie^{ik_{2}}\\
0 & 0 & 0 & \cdots & ie^{ik_{2}} & 0\\
0 & 0 & ie^{ik_{2}} & \cdots & 0 & 0\\
0 & ie^{ik_{2}} & 0 & \cdots & 0 & 0
\end{pmatrix}.
\end{equation}
This is a unitary matrix satisfying the following relation with the Hamiltonian 
\begin{equation}
{\cal R}_{k_{2}}^{\dagger}{\cal H}_{k_{1},k_{2}}{\cal R}_{k_{2}}={\cal H}_{k_{1},-k_{2}}.\label{eq:Rham}
\end{equation}
Despite the dependence on $k_{2}$ the eigenvectors of ${\cal R}_{k_{2}}$
can be found as not depending on $k_{2}$ and its diagonal form is
given by the equation, 
\begin{equation}
{\cal U}^{\dagger}{\cal R}_{k_{2}}{\cal U}=\begin{pmatrix}\mathbf{1}_{2} & {\bf 0} & {\bf 0} & {\bf 0}\\
{\bf 0} & e^{ik_{2}}\mathbf{1}_{2N_{\downarrow}-2} & {\bf 0} & {\bf 0}\\
{\bf 0} & {\bf 0} & -\mathbf{1}_{2} & {\bf 0}\\
{\bf 0} & {\bf 0} & {\bf 0} & -e^{ik_{2}}\mathbf{1}_{2N_{\downarrow}-2}
\end{pmatrix},
\end{equation}
where ${\cal U}$ is the eigenbasis and the blocks are the unity matrices
of size $2$ or $2N_{\downarrow}-2$. We note that for any zig-zag
segment length there are two eigenvalues of amplitude $2$, and other two with value $-2$ while the remaining part of the spectrum is
given by $\pm e^{ik_{2}}$. The spectrum is chiral in the sense that for
any eigenvalue there is a partner with opposite energy. In the reflection
planes $k_{2}=0,\pi$ the reflection operator becomes a symmetry for
the Hamiltonian, i.e.
\begin{equation}
\left[{\cal H}_{k_{1},0(\pi)},{\cal R}_{0(\pi)}\right]=0
\end{equation}
and the $k$-dependence in the spectrum of ${\cal R}_{k_{2}}$ vanishes, being equal to $+1$
at $k_{2}=0$ and $-1$ at $k_{2}=\pi$.

A less straightforward spatial symmetry is that provided by the non-symmorphic glide transformation. 
It is obtained from the product
of a normal reflection with a reflection plane being perpendicular to the
one involved in ${\cal R}_{k_{2}}$ and a translation $\vec{t}$ in
a direction parallel to the reflection plane. One can easily find
that for any zig-zag segment length it is always given by $\vec{t}=\frac{1}{2}\vec{a}_{2}$.
The action of the glide operation is shown in Fig.\ref{fig:patt_PDs}(a)-(b).
We observe that for any zig-zag with even $L_{z}$ both the reflection and
translation of the non-primitive lattice vector do not map the original square lattice
into itself, only their product does it. Similarly to the
regular reflection the glide symmetry is given by a unitary matrix that is mixing the
$a$ and $b$ orbitals with a block structure,
\begin{equation}
{\cal R}_{k_{1},k_{2}}^{t}=\begin{pmatrix}{\bf 0} & {\bf 0} & -{\bf R}_{k_{1},k_{2}}^{\mathbf{t}} & {\bf 0}\\
{\bf 0} & {\bf 0} & {\bf 0} & -e^{ik}{\bf {\bf R}}_{k_{1},k_{2}}^{\mathbf{t}}\\
{\bf {\bf R}}_{k_{1},k_{2}}^{\mathbf{t}} & {\bf 0} & {\bf 0} & {\bf 0}\\
{\bf 0} & e^{ik}{\bf {\bf R}}_{k_{1},k_{2}}^{\mathbf{t}} & {\bf 0} & {\bf 0}
\end{pmatrix},
\end{equation}
and the blocks given by the sub-blocks of the size $L_{z}-1\equiv N_{\downarrow}/2$,
\begin{equation}
{\bf {\bf R}}_{k_{1},k_{2}}^{\mathbf{t}}=\begin{pmatrix}\mathbf{0} & ie^{-i\frac{k_{2}}{2}}\mathbf{1}_{L_{z}-1}\\
ie^{i\frac{k_{2}}{2}}\mathbf{1}_{L_{z}-1} & \mathbf{0}
\end{pmatrix}.
\end{equation}
${\cal R}_{k_{1},k_{2}}^{t}$ carries the intrinsic dependence on
both quasimomenta that is a direct consequence of the non-symmorphic nature of the
glide. One cannot find a unit cell that would map onto itself under
such an operation. The relation with the Hamiltonian is the same as
for a normal reflection mapping $k_{1}$ into $-k_{1}$, i.e.,

\begin{equation}
{\cal R}_{k_{1},k_{2}}^{t\dagger}{\cal H}_{k_{1},k_{2}}{\cal R}_{k_{1},k_{2}}^{t}={\cal H}_{-k_{1},k_{2}}.\label{eq:Rtham}
\end{equation}
The glide operator ${\cal R}_{k_{1},k_{2}}^{t}$ has eigenvectors
that depend only on $k_{2}$ and eigenvalues that depend only on $k_{1}$,
thus its diagonal form is given by the equation 
\begin{equation}
{\cal V}_{k_{2}}^{\dagger}{\cal R}_{k_{1},k_{2}}^{t}{\cal V}_{k_{2}}=\begin{pmatrix}{\bf 1}_{N_{\downarrow}} & {\bf 0} & {\bf 0} & {\bf 0}\\
{\bf 0} & e^{ik_{1}}{\bf {\bf 1}}_{N_{\downarrow}} & {\bf 0} & {\bf 0}\\
{\bf 0} & {\bf 0} & -{\bf 1}_{N_{\downarrow}} & {\bf 0}\\
{\bf 0} & {\bf 0} & {\bf 0} & -e^{ik_{1}}{\bf {\bf 1}}_{N_{\downarrow}}
\end{pmatrix},
\end{equation}
where ${\cal V}_{k_{2}}$ is the eigenbasis and the blocks are the
unity matrices of the size $N_{\downarrow}$. In the glide planes
$k_1=0,\pi$ the glide operator becomes a symmetry of the Hamiltonian,
i.e.
\begin{equation}
\left[{\cal H}_{0(\pi),k_{2}},{\cal R}_{0(\pi),k_{2}}^{t}\right]=0,
\end{equation}
and the $k$-dependent eigenvalues change the sign when going from
one glide plane to the other. 

Having two reflection-like operators, as given by Eqs. (\ref{eq:Rham})
and Eq. (\ref{eq:Rtham}), one can immediately construct an inversion
operator ${\cal I}_{k_{1},k_{2}}$ satisfying, 
\begin{equation}
{\cal I}_{k_{1},k_{2}}^{\dagger}{\cal H}_{k_{1},k_{2}}{\cal I}_{k_{1},k_{2}}={\cal H}_{-k_{1},-k_{2}}.\label{eq:Iham}
\end{equation}
Looking at Eqs. (\ref{eq:Rham}) and (\ref{eq:Rtham}) we argue that the inversion operator has the form 
\begin{equation}
{\cal I}_{k_{1},k_{2}}\equiv-e^{-i\frac{k_{2}}{2}}{\cal R}_{k_{2}}{\cal R}_{k_{1},-k_{2}}^{t},
\end{equation}
where the phase factor is chosen only for convenience.
We note that the operators in the product are taken at opposite $k_{2}$
because inserting ${\cal I}_{k_{1},k_{2}}$ into Eq. (\ref{eq:Iham})
the first action is on the Hamiltonian with ${\cal R}_{k_{2}}$ which gives us
${\cal H}_{k_{1},-k_{2}}$. Then, if we want to connect with ${\cal H}_{-k_{1},-k_{2}}$
we have to use glide operator at point $(k_{1},-k_{2})$. Unlike the
reflection and glide, the inversion does not mix the orbital sectors
and its block structure is 
\begin{equation}
{\cal I}_{k_{1},k_{2}}=\begin{pmatrix}\mathbf{I}_{k_{2}} & {\bf 0} & {\bf 0} & {\bf 0}\\
{\bf 0} & e^{ik}\mathbf{I}_{k_{2}} & {\bf 0} & {\bf 0}\\
{\bf 0} & {\bf 0} & \mathbf{I}_{k_{2}} & {\bf 0}\\
{\bf 0} & {\bf 0} & {\bf 0} & e^{ik}\mathbf{I}_{k_{2}}
\end{pmatrix},
\end{equation}
with $\mathbf{I}_{k_{2}}$ defined bythe  diagonal sub-blocks of the size
$L_{z}$ and $L_{z}-2$ 
\begin{equation}
\mathbf{I}_{k_{2}}=\begin{pmatrix}\mathbf{P}_{L_{z}} & \mathbf{0}\\
\mathbf{0} & e^{ik_{2}}\mathbf{P}_{L_{z}-2}
\end{pmatrix}.
\end{equation}
These blocks correspond to the vertical and horizontal sections of
the spin down/up segment in the unit cell, being the sites $i=1,2,3,4$
and $i=5,6$ for $L_{z}=4$ unit cell as shown in Fig. \ref{fig:cell}.
Finally, ${\bf P}_{n}$ is a simple reflection operator for these straight
sections of a zig-zag pattern given by $n\times n$ antidiagonal matrix, 
\begin{equation}
\mathbf{P}_{n}=\begin{pmatrix}0 & 0 & \cdots & 0 & 1\\
0 & 0 & \cdots & 1 & 0\\
0 & 1 & \cdots & 0 & 0\\
1 & 0 & \cdots & 0 & 0
\end{pmatrix}.
\end{equation}
Note that any multiplication of a glide, or a reflection with a translation,
with an ordinary reflection, the resulting inversion does not contain
a dependence on a translation vector. Indeed, looking at Fig. \ref{fig:patt_PDs}(a)-(b) we can easily
find inversion centers for zig-zags with $L_{z}=2$ and $L_{z}=3$.
The $k$-dependence in ${\cal I}_{k_{1},k_{2}}$ comes from the fact
that the selected unit cell does not map onto itself under the inversion.
Hence, one can easily check that it is not possible to find a
unit cell which is compatible with inversion for any segment length
$L_{z}$. The case of even and odd $L_{z}$ are qualitatively different
because for even $L_{z}$ the inversion center does not coincide with any
lattice site whereas for odd $L_{z}$ it is a central site in any
vertical or horizontal section of a zig-zag. For this reason the spectrum
of ${\cal I}_{k_{1},k_{2}}$ for zig-zags with even segment length
$L_{z}$ is different than for odd $L_{z}$. For even $L_{z}$ we
have a chiral spectrum of the form, 
\begin{eqnarray}
\left\{ \left\{ 1\right\} ^{L_{z}}\!\!,\left\{ -1\right\} ^{L_{z}}\!\!,\left\{ e^{ik_{1}}\right\} ^{L_{z}}\!\!,\left\{ -e^{ik_{1}}\right\} ^{L_{z}}\!\!,\left\{ e^{ik_{2}}\right\} ^{L_{z}}\!\!,\right.\nonumber \\
\left.\left\{ -e^{ik_{2}}\right\} ^{L_{z}}\!\!,\left\{ e^{i(k_{1}\!+\! k_{2})}\right\} ^{L_{z}\!-\!2}\!\!,\left\{ -e^{i(k_{1}\!+\! k_{2})}\right\} ^{L_{z}\!-\!2}\right\} \!,\quad\nonumber
\end{eqnarray}
where the notation means that, e.g., the eingevalue $1$ is $L_{z}$-fold.
For odd $L_{z}$ the spectrum has different number of positive and
negative entries, i.e. 
\begin{eqnarray}
\left\{ \left\{ 1\right\} ^{L_{z}\!+\!1}\!\!,\left\{ -1\right\} ^{L_{z}\!-\!1}\!\!,\left\{ e^{ik_{1}}\right\} ^{L_{z}\!+\!1}\!\!,\left\{ -e^{ik_{1}}\right\} ^{L_{z}\!-\!1}\!\!,\left\{ e^{ik_{2}}\right\} ^{L_{z}\!-\!1}\!,\right.\nonumber \\
\left.\left\{ -e^{ik_{2}}\right\} ^{L_{z}\!-\!3}\!\!,\left\{ e^{i(k_{1}\!+\! k_{2})}\right\} ^{L_{z}\!-\!1}\!\!,\left\{ -e^{i(k_{1}\!+\! k_{2})}\right\} ^{L_{z}\!-\!3}\right\} \!.\quad\quad\nonumber
\end{eqnarray}
For odd $L_{z}$ we can always put electrons in the inversion centers
that coincide with physical sites of the system. There are four inversion
centers in each unit cell and two orbitals per site so in this way
we can obtain eight eigenstates of ${\cal I}_{k_{1},k_{2}}$ with
{\it positive} eigenvalues, $\left\{ 1,e^{ik_{1}},e^{ik_{2}},e^{i(k_{1}+k_{2})}\right\} $,
all of them being double degenerate. This observation explains why there are more {\it positive}
eigenvalues then {\it negative} for the case of odd $L_{z}$. Note that
despite the $k$-dependence in the eigenvalues, the eigenvectors of
${\cal I}_{k_{1},k_{2}}$ are $k$-independent for any $L_{z}$.

\subsection{Commutation relations of spacial and non-spatial symmetries\label{sec:commu}}

It is not obvious at first glance that the reflection and the glide operators
commute. Calculating their commutator one has to always bear in mind
that it is the relation to the Hamiltonian that matters, so it is
crucial to update the $\vec{k}$ points for which we consider the symmetry
operators. For instance, one can show that the operation of reflection
and glide does not depend from their order with respect to the Hamiltonian which means
that
\begin{equation}
{\cal R}_{k_{2}}{\cal R}_{k_{1},-k_{2}}^{t}-{\cal R}_{k_{1},k_{2}}^{t}{\cal R}_{k_{2}}\equiv0.
\end{equation}
Thus, similarly one can demonstrate that the inversion operator commutes with the above
operators. Concerning the relation with respect to the time reversal operation
one finds that both normal reflection and glide commute with ${\cal T}$, that is
\begin{equation}
{\cal R}_{k_{2}}{\cal T}-{\cal T}{\cal R}_{-k_{2}}^{\star}\equiv0,
\end{equation}
and
\begin{equation}
{\cal R}_{k_{1},k_{2}}^{t}{\cal T}-{\cal T}{\cal R}_{-k_{1},-k_{2}}^{t\star}\equiv0,
\end{equation}
where the symbol star indicate complex conjugation. Thus, we conclude that the same property holds for
the inversion. Finally, concerning the chirality (sublattice) symmetry
we find that the reflection commutes with ${\cal S}_{k_{1}}$, 
\begin{equation}
{\cal R}_{k_{2}}{\cal S}_{k_{1}}-{\cal S}_{k_{1}}{\cal R}_{k_{2}}\equiv0,\label{eq:comRS}
\end{equation}
and the glide anticommutes or commutes with ${\cal S}_{k_{1}}$ for
even or odd $L_{z}$ respectively, i.e.
\begin{equation}
{\cal R}_{k_{1},k_{2}}^{t}{\cal S}_{k_{1}}+(-1)^{L_{z}}{\cal S}_{k_{1}}{\cal R}_{k_{1},k_{2}}^{t}\equiv0.\label{eq:comRtS}
\end{equation}
Intuitively one can argue that the chirality symmetry is related to
the two-sublattice structure of the problem. Since under reflection
a site belonging to one sublattice transforms to another site from
the same sublattice we get the vanishing commutator in (\ref{eq:comRS}).
On the other hand, looking at Fig. \ref{fig:cell}, we notice that
for $L_{z}=2$ the glide does mix the sites from different sublattices
hence one gets a vanishing anticommutator. It is easy to check that this
happens for any even $L_{z}$ whereas for the odd ones we always get a vanishing
commutation relation. Finally, we note that since all spacial symmetries commute
with time reversal, their commutation relation with particle-hole
symmetry is the same as that one with the chirality.

\subsection{Combined symmetries\label{sec:combi}}

\begin{figure}
\includegraphics[clip,width=0.5\columnwidth]{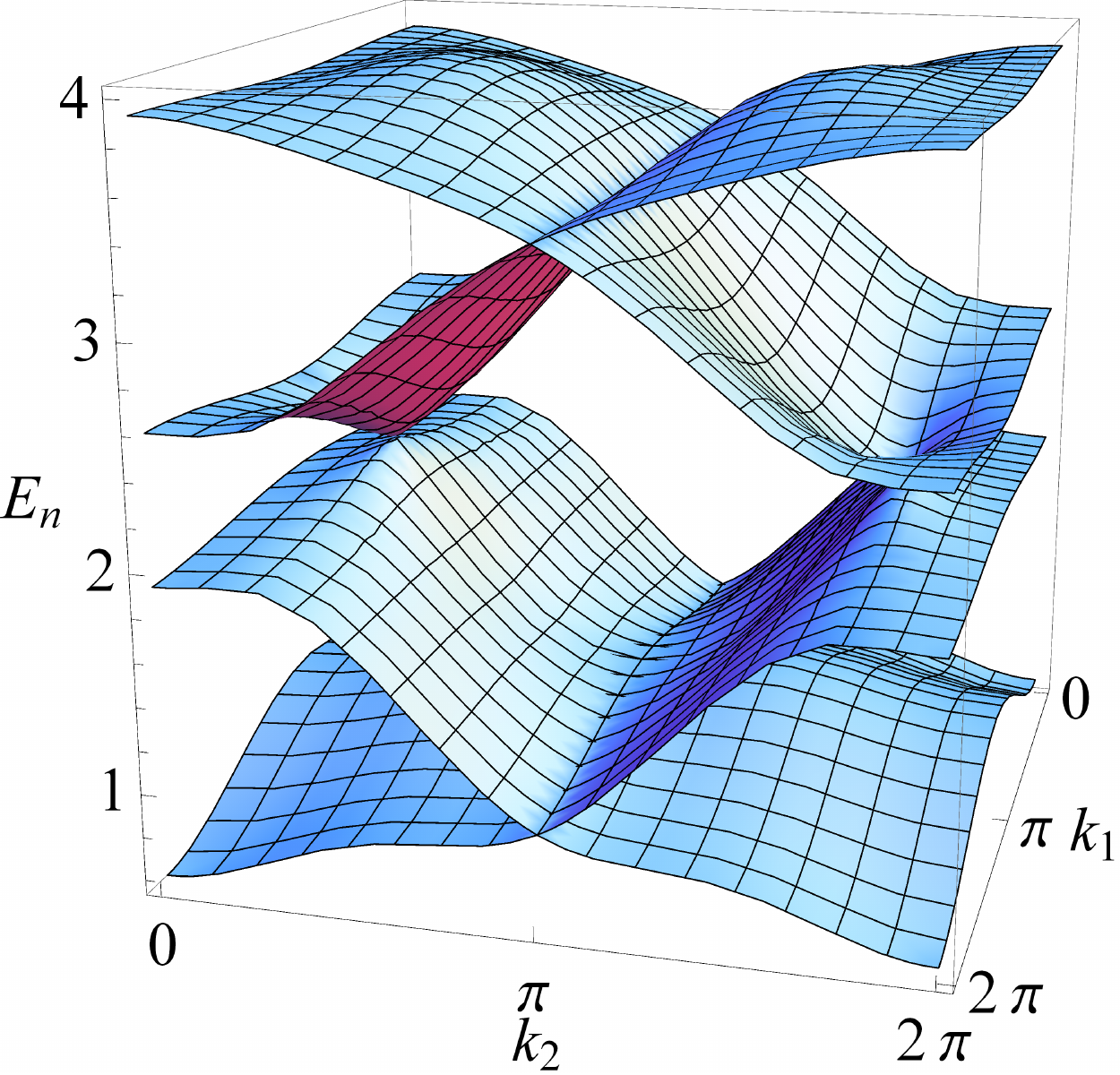}
\includegraphics[clip,width=0.5\columnwidth]{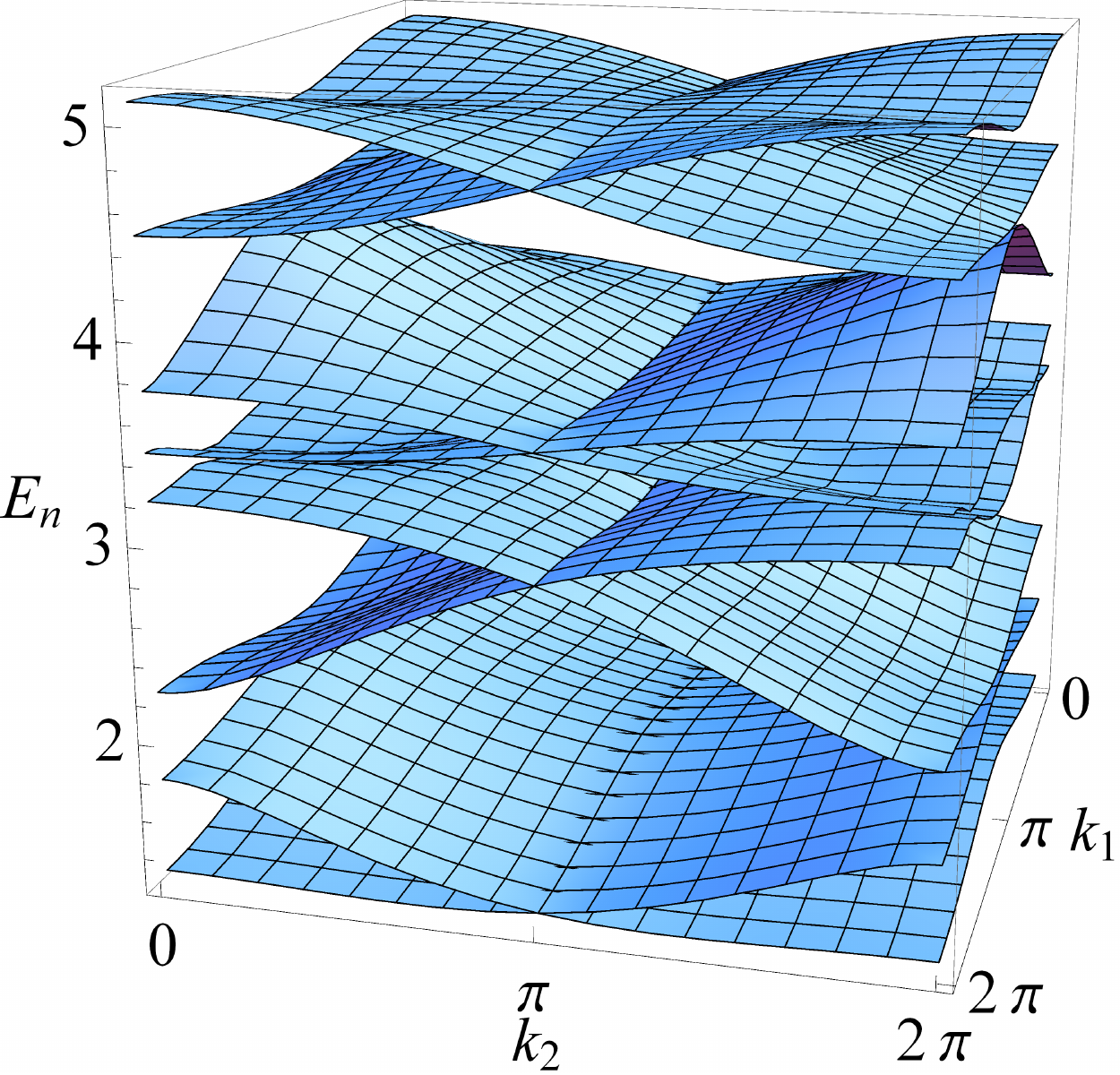}
\protect\caption{Positive-energy bands of the undistorted (cubic) zig-zag $L_{z}=2$
(left) and $L_{z}=3$ (right) systems. The negative-energy bands can
be recovered by reflection with respect to zero energy due to the chirality
${\cal S}_{k_{1}}$. We note that the bands appear in $(2L_{z}-2)$ pairs
and every pair has a 1D crossing line at $k_{2}=\pi$ due to
effective Kramers degeneracy occurring at any cut along $k_{2}$ direction. These pairs
are Kramers doublets at $k_{2}=\pi$. The spectrum for $L_{z}=2$ has
two Dirac points in the glide plane $k_{1}=\pi$. \label{fig:specs}}
\end{figure}
Having both spatial and non-spatial symmetries we can construct various
combined symmetries by taking their products. Here, we will focus only
on antiunitary symmetries constructed by combining time reversal
operator with reflection, glide and inversion. We start with the reflection,
defining ${\cal T}_{k_{2}}^{(1)}$ as 
\begin{equation}
{\cal T}_{k_{2}}^{(1)}\equiv{\cal R}_{k_{2}}{\cal T},
\end{equation}
we obtain a time reversal operator acting in the 1D cuts of BZ parallel
to the $k_{1}$ ($k_{1}$-cuts) axis, 
\begin{equation}
{\cal T}_{k_{2}}^{(1)\dagger}{\cal H}_{k_{1},k_{2}}{\cal T}_{k_{2}}^{(1)}={\cal H}_{-k_{1},k_{2}}^{T}.
\end{equation}
In the same way using glide operator we can define ${\cal T}_{k_{1},k_{2}}^{(2)}$
as, 
\begin{equation}
{\cal T}_{k_{1},k_{2}}^{(2)}\equiv{\cal R}_{k_{1},k_{2}}^{t}{\cal T},
\end{equation}
which is a time reversal operator acting in the cuts parallel to the
$k_{2}$ axis ($k_{2}$-cuts), 

\begin{equation}
{\cal T}_{k_{1},k_{2}}^{(2)\dagger}{\cal H}_{k_{1},k_{2}}{\cal T}_{k_{1},k_{2}}^{(2)}={\cal H}_{k_{1},-k_{2}}^{T}.
\end{equation}
Similarly, we can combine time reversal with the inversion 
operator by introducing the operator ${\cal K}_{k_{1},k_{2}}$ as given by
\begin{equation}
{\cal K}_{k_{1},k_{2}}\equiv{\cal I}_{k_{1},k_{2}}{\cal T},
\end{equation}
whose action on the Hamiltonian yields its transpose or complex conjugate 
\begin{equation}
{\cal K}_{k_{1},k_{2}}^{\dagger}{\cal H}_{k_{1},k_{2}}{\cal K}_{k_{1},k_{2}}={\cal H}_{k_{1},k_{2}}^{T}.
\end{equation}
Thus, we can denote ${\cal K}_{k_{1},k_{2}}$ as a {\it pure conjugation}
operator. A close inspection at the powers of these three anti-unitary
symmetries, for $k_{1}$-cuts in the Brillouin zone, and for the pure conjugation, we find that these
square to give the unity
\begin{equation}
{\cal T}_{k_{2}}^{(1)}{\cal T}_{k_{2}}^{(1)\star}\equiv1,\quad{\cal K}_{k_{1},k_{2}}{\cal K}_{k_{1},k_{2}}^{\star}\equiv1,
\end{equation}
on the other hand, for $k_{2}$-cuts we obtain a $k$-dependent result,
\begin{equation}
{\cal T}_{k_{1},k_{2}}^{(2)}{\cal T}_{k_{1},-k_{2}}^{(2)\star}\equiv e^{-ik_{2}}.
\end{equation}
This means that in two time-reversal points of a given $k_{2}$-cut,
namely $k_{2}=0$ and $k_{2}=\pi$ we have $({\cal T}^{(2)})^{2}=1$
and $({\cal T}^{(2)})^{2}=-1$. Thus, one can conclude that for every $k_{2}$-cut
the point $k_{2}=\pi$ is a Kramers degenerate point. Note that this
is a surprising result as the problem with which we deal is spinless
and also because the other time-reversal point in the Brillouin zone is not Kramers degenerate.
The result on the bands of any $L_{z}$ zig-zag system is interesting as
at $k_{2}=\pi$ and any $k_{1}$ every band is twice degenerate.
To explicitly demonstrate this property, we present the 1D bands crossings for $L_{z}=2$ and $L_{z}=3$
in Fig. \ref{fig:specs}.

\subsection{Gauging away $k$-dependecies in all symmetries operators\label{sec:gaug_all}}

As we noticed in the previous Sections, due to the structure of 
the unit cell and the presence of the non-symmorphic glide symmetry, most of the spatial
and non-spatial symmetries are $k$-dependent. It is worth and interesting to underline that
these $k$-dependences can be simultaneously gauged
away for any zig-zag segment length $L_{z}$ by a transformation
of the form
\begin{equation}
{\cal G}_{\vec{k}}=\begin{pmatrix}{\bf G}_{\downarrow} & {\bf 0} & {\bf 0} & {\bf 0}\\
{\bf 0} & {\bf G}_{\uparrow} & {\bf 0} & {\bf 0}\\
{\bf 0} & {\bf 0} & {\bf G}_{\downarrow} & {\bf 0}\\
{\bf 0} & {\bf 0} & {\bf 0} & {\bf G}_{\uparrow}
\end{pmatrix},
\end{equation}
with diagonal blocks ${\bf G}_{\downarrow}$and ${\bf G}_{\uparrow}$
of the size $N_{\downarrow}=N_{\uparrow}=2L_{z}-2$, 
\begin{equation}
{\bf G}_{\downarrow}=e^{-ik_{1}}\begin{pmatrix}1 & 0 & 0 & \cdots & 0\\
0 & e^{i\frac{1}{N_{\downarrow}}k_{2}} & 0 & \cdots & 0\\
0 & 0 & e^{i\frac{2}{N_{\downarrow}}k_{2}} & \cdots & 0\\
\vdots & \vdots & \vdots & \ddots & 0\\
0 & 0 & 0 & 0 & e^{i\frac{N_{\downarrow}-1}{N_{\downarrow}}k_{2}}
\end{pmatrix},
\end{equation}
and
\begin{equation}
{\bf G}_{\uparrow}=e^{i\frac{k_{1}}{2}}{\bf G}_{\downarrow},
\end{equation}
corresponding to the domains of spin up/down in the unit call. Note
that these matrices are the same for the orbital $a$ and orbital $b$
sector. The Hamiltonian transforms as a linear operator under the
basis rotation, i.e.
\begin{equation}
{\cal \tilde{H}}_{\vec{k}}\equiv{\cal G}_{\vec{k}}^{\dagger}{\cal H}_{\vec{k}}{\cal G}_{\vec{k}},
\end{equation}
where tilde indicates the operator in the gauge transformed basis. Now, if we require
that the reflection operator in the gauge transformed basis, i.e., ${\cal \tilde{R}}$
acts as a reflection operator with respect to ${\cal \tilde{H}}_{\vec{k}}$,
i.e., 
\begin{equation}
\tilde{{\cal R}}{}^{\dagger}\tilde{{\cal H}}_{k_{1},k_{2}}\tilde{{\cal R}}=\tilde{{\cal H}}_{k_{1},-k_{2}},
\end{equation}
we easily find that ${\cal \tilde{R}}$ should have the following form 
\begin{equation}
{\cal \tilde{R}}\propto{\cal G}_{k_{1},k_{2}}^{\dagger}{\cal R}_{k_{2}}{\cal G}_{k_{1},-k_{2}},
\end{equation}
where the complex prefactor can be chosen in such a way to completely
remove any $k$-dependence in ${\cal \tilde{R}}$. Similarly, we can
proceed for the glide, 
\begin{equation}
{\cal \tilde{R}}^{t}\propto{\cal G}_{k_{1},k_{2}}^{\dagger}{\cal R}_{k_{1},k_{2}}^{t}{\cal G}_{-k_{1},k_{2}},
\end{equation}
and for the other non-spatial symmetries. For instance time reversal operator,
\begin{equation}
{\cal \tilde{T}}\propto{\cal G}_{k_{1},k_{2}}^{\dagger}{\cal T}{\cal G}_{-k_{1},-k_{2}}^{\star},
\end{equation}
where the gauge matrix on the right is taken with complex conjugate.
While for the chirality we have that
\begin{equation}
{\cal \tilde{S}}\propto{\cal G}_{k_{1},k_{2}}^{\dagger}{\cal S}_{k_{2}}{\cal G}_{k_{1},k_{2}},
\end{equation}
and transforms through a simple basis rotation. The rest of the symmetries
can be constructed by taking the product of the above operators, i.e.,
\begin{equation}
\tilde{{\cal I}}={\cal \tilde{R}}{\cal \tilde{R}}^{t}={\cal \tilde{R}}^{t}{\cal \tilde{R}},
\end{equation}
to get the inversion and
\begin{equation}
\tilde{{\cal C}}={\cal \tilde{T}}{\cal \tilde{S}}^{\star}={\cal \tilde{S}}{\cal \tilde{T}},
\end{equation}
to get charge conjugation. Note that the spatial symmetries and the
time reversal do not transform as linear operators under basis rotation.
For this reason their spectra are different in the gauge transformed basis with
respect to the initial one and their commutation relations could be
different as well. Accidentally, we find that the commutation relations
remain the same as shown in previous Section. 

Finally, we stress that the consequence of the removal of the $k$-dependencies
from the symmetries operators is to alter the periodicity of the Hamiltonian in the momentum space.
One find that for a given $L_{z}$ the period in $k_{1}$ is always
doubled, i.e., $\tilde{{\cal H}}_{k_{1}+4\pi,k_{2}}=\tilde{{\cal H}}_{k_{1},k_{2}}$
whereas that one in $k_{2}$ is increased $N_{\downarrow}$ - times. This
effect can be captured by what we call the {\it shift} operator. We
find that however the period is elongated the old period of $2\pi$
survives up to a basis rotation described by a unitary shift operator
$\chi$. For $k_{1}$ we find that, 
\begin{equation}
\tilde{{\cal H}}_{k_{1}+2\pi,k_{2}}=\chi_{1}^{\dagger}\tilde{{\cal H}}_{k_{1},k_{2}}\chi_{1},\label{eq:hamsh1}
\end{equation}
where $\chi_{1}$ can be found as a '$2\pi$-basis mismatch' of the
gauge matrix, i.e., 
\begin{equation}
\chi_{1}\propto{\cal G}_{k_{1},k_{2}}^{\dagger}{\cal G}_{k_{1}+2\pi,k_{2}}.
\end{equation}
One finds that 
\begin{equation}
\left(\chi_{1}\right)^{2}=1,
\end{equation}
meaning that after two $2\pi$-shifts in $k_{1}$ we recover the same
Hamiltonian $\tilde{{\cal H}}_{k_{1},k_{2}}$. Similarly for $k_{2}$
we observe, 
\begin{equation}
\tilde{{\cal H}}_{k_{1},k_{2}+2\pi}=\chi_{2}^{\dagger}\tilde{{\cal H}}_{k_{1},k_{2}}\chi_{2},\label{eq:hamsh2}
\end{equation}
with $\chi_{2}$ unitary shift operator having the form of 
\begin{equation}
\chi_{2}\propto{\cal G}_{k_{1},k_{2}}^{\dagger}{\cal G}_{k_{1},k_{2}+2\pi},
\end{equation}
and becoming unity after $N_{\downarrow}$ applications 
\begin{equation}
\left(\chi_{2}\right)^{N_{\downarrow}}=1.
\end{equation}

\subsection{Gauging away $k$-dependence of the reflection operator\label{sec:gaug_refl}}

The gauge transformation described in Sec. \ref{sec:gaug_all} makes
all the symmetry opearators $k$-independent but concomitantly one has to
increase the effective Brillouin zone of the Hamiltonian. This is caused by the
fact that the unit cell is not left invariant by any of the symmetries
of the whole system. In the case of glide, inversion and chirality symmetries
it is in indeed impossible to define a unit cell that would map onto
itself under their action. However, in the case of reflection this transformation is
possible. Another operating scheme would be to choose a square unit cell, as shown
in Fig. \ref{fig:patt_PDs}(a)-(b), but in this way we increase the dimensionality
of the operators and create artificial symmetries coming from the
multiple copies of the elementary cell. The other solution, which
is much more convenient, is to slightly modify the elementary unit
cell shown in Fig. \ref{fig:cell}. One has to remind that every
physical site in the lattice has two orbital flavors so the resulting system
can be also mapped into an effective bilayer. The reflection can be seen as a $\pi$-rotation
of the bilayer with axis along the $\vec{a}_{1}$ direction. Thus
a reflection-invariant unit cell can be constructed by taking the
original cell for orbital $b$ and for orbital $a$ one has to move
the first two sites ($i=1,7$ in Fig. \ref{fig:cell}) of the vertical
segments after the last two sites ($i=6,12$ in Fig. \ref{fig:cell})
of the horizontal segments. This can be realized by a gauge
transformation of the type
\begin{equation}
{\cal G}_{k_{2}}^{{\cal R}}=\begin{pmatrix}{\bf 1}_{\downarrow}^b & {\bf 0} & {\bf 0} & {\bf 0}\\
{\bf 0} & {\bf 1}_{\uparrow}^b & {\bf 0} & {\bf 0}\\
{\bf 0} & {\bf 0} & {\bf G}_{\downarrow}^{R} & {\bf 0}\\
{\bf 0} & {\bf 0} & {\bf 0} & {\bf G}_{\uparrow}^{R}
\end{pmatrix},
\end{equation}
with two identical sub-blocks for orbital $a$ segment,
\begin{equation}
{\bf G}_{\downarrow,\uparrow}^{R}=\begin{pmatrix}e^{-ik_{2}} & 0 & 0 & \cdots & 0\\
0 & 1 & 0 & \cdots & 0\\
0 & 0 & 1 & \cdots & 0\\
\vdots & \vdots & \vdots & \ddots & 0\\
0 & 0 & 0 & 0 & 1
\end{pmatrix},
\end{equation}
carrying the gauge for the first sites of the vertical segment. After
the covariant transformation of the reflection operator, i.e., ${\cal R}_{k_{2}}\to{\cal G}_{k_{2}}^{{\cal R}\dagger}{\cal R}_{k_{2}}{\cal G}_{-k_{2}}^{{\cal R}}$
and extracting global phase factor we find it completely $k$-independent.

\subsection{Gauging away $k_1$-dependence of the glide symmetry operator\label{sec:gaug_refl2}}

Another gauge transformation can be found to demonstrate that the glide operator
${\cal R}_{\vec{k}}^{t}$ is dependent only on $k_{2}$ without
affecting the periodicity of the Hamiltonian. Again, this solution
is equivalent to a modification of the elementary unit cell
shown in Fig. \ref{fig:cell}. Treating the $a$ and $b$ orbital
degrees of freedom as two layers one can treat the glide as a $\pi$-rotation
of the bilayer with axis along the $\vec{a}_{2}$ direction followed
by a shift of $\vec{a}_{2}/2$ . Thus, a way to construct a unit cell
which is mostly compatible with a glide is to take the original cell
for the orbital $a$ and for the orbital $b$, and then shift the spin down domain 
by $\vec{a}_{1}$ (see Fig. \ref{fig:cell}). In this way
we obtain a gauge transformation of the form
\begin{equation}
{\cal G}_{k_{1}}^{{\cal R}^{t}}=\begin{pmatrix}{\bf G}_{\downarrow}^{R^{t}} & {\bf 0} & {\bf 0} & {\bf 0}\\
{\bf 0} & {\bf 1}_{\uparrow}^b & {\bf 0} & {\bf 0}\\
{\bf 0} & {\bf 0} & {\bf 1}_{\downarrow}^a & {\bf 0}\\
{\bf 0} & {\bf 0} & {\bf 0} & {\bf 1}_{\uparrow}^a
\end{pmatrix},
\end{equation}
with one non-trivial subblock,
\begin{equation}
{\bf G}_{\downarrow}^{R^{t}}=\begin{pmatrix}e^{-ik_{1}} & 0 & 0 & \cdots & 0\\
0 & 1 & 0 & \cdots & 0\\
0 & 0 & 1 & \cdots & 0\\
\vdots & \vdots & \vdots & \ddots & 0\\
0 & 0 & 0 & 0 & 1
\end{pmatrix},
\end{equation}
carrying the gauge for the spin down domain of orbitals $b$. After
the covariant transformation of the glide operator, i.e., ${\cal R}_{\vec{k}}^{t}\to{\cal G}_{k_{1}}^{{\cal R}^{t}\dagger}{\cal R}_{\vec{k}}^{t}{\cal G}_{-k_{1}}^{{\cal R}^{t}}$
and extracting global phase factor we find it dependent only on $k_{2}/2$
(due to the $\vec{a}_{2}/2$ shift) in such a way that it eigenvalues
are $k$-independent taking values $g_{\pm}=\pm1$. The dependence on $k_2$ of the eigenbasis
is such that the eigenvectors with $g_+$ eigenvalues at fixed $k_2$ become the $g_-$ eigenvectors
at $k_2+2\pi$.

\subsection{Hamiltonians and k-independent symmetries for $L_{z}=2,3$\label{sec:hams=000026syms_gaug}}

We will focus on undistorted zig-zag systems where the only non-zero
hopping amplitudes are $t_{\hat{a},i}^{bb}=t_{\hat{b},i}^{aa}=-t$
and where we set $t=1$ as the energy unit. For zig-zag segment lengths
$L_{z}=2$ and $L_{z}=3$ the Hamiltonians are $8\times8$ and $16\times16$
hermitian matrices, respectively. Thus, it is convenient to use the
product basis of $3$ and $4$ Pauli matrices to show the exact form
of the Hamiltonians and their symmetries in these two cases. For $L_{z}=2$
we will decompose the operators in basis of $64$ hermitian matrices
whose building blocks are pseudospins $S=1/2$ defined on three {\it artificial}
sites, i.e.
\begin{equation}
\sigma_{1}^{\alpha}=\sigma^{\alpha}\otimes1_{2}\otimes1_{2},\quad\sigma_{2}^{\alpha}=1_{2}\otimes\sigma^{\alpha}\otimes1_{2},\quad\sigma_{3}^{\alpha}=1_{2}\otimes1_{2}\otimes\sigma^{\alpha},\label{eq:paul3}
\end{equation}
with $\alpha=x,y,z$ and $\sigma^{\alpha}$ being a Pauli matrix.
Hence, the Hamiltonian in the gauge transformed basis and for the shortest possible
zig-zag with $L_{z}=2$ can be represented as
\begin{eqnarray}
\tilde{{\cal H}}_{\vec{k}} & = & \sin\frac{k_{2}}{2}\left(\sin\frac{k_{1}}{2}\sigma_{3}^{x}+\cos\frac{k_{1}}{2}\sigma_{3}^{y}\right)\sigma_{1}^{z}\sigma_{2}^{x}\nonumber \\
 & + & \cos\frac{k_{2}}{2}\left(\sin\frac{k_{1}}{2}\sigma_{3}^{x}-\cos\frac{k_{1}}{2}\sigma_{3}^{x}\right)\sigma_{2}^{x}\nonumber \\
 & - & \cos\frac{k_{2}}{2}\sigma_{3}^{x}-\sin\frac{k_{2}}{2}\sigma_{1}^{z}\sigma_{3}^{y}\nonumber \\
 & + & J_{H}\sigma_{2}^{z}-\lambda\sigma_{1}^{y}\sigma_{2}^{z}.
\end{eqnarray}
The spatial symmetries take the form
\begin{eqnarray}
{\cal \tilde{R}} & = & \sigma_{1}^{y},\nonumber \\
{\cal \tilde{R}}^{t} & = & \sigma_{1}^{y}\sigma_{3}^{x},\nonumber \\
\tilde{{\cal I}} & = & \sigma_{3}^{x},
\end{eqnarray}
and the non-spatial symmetries are
\begin{eqnarray}
{\cal \tilde{T}} & = & \sigma_{1}^{z},\nonumber \\
{\cal \tilde{S}} & = & \sigma_{2}^{x}\sigma_{3}^{z},\nonumber \\
\tilde{C} & = & \sigma_{1}^{z}\sigma_{2}^{x}\sigma_{3}^{z}.
\end{eqnarray}
The algebra is completed by the shift operators defined in the previous
subsection as
\begin{eqnarray}
\chi_{1} & = & \sigma_{2}^{z},\nonumber \\
\chi_{2} & = & \sigma_{3}^{z}.
\end{eqnarray}
It is instructive to check that all these operators really satisfy the
relevant relations with respect to the Hamiltonian. 

Analogically, for $L_{z}=3$ we span a basis of $256$ hermitian matrices
whose building blocks are pseudospins $S=1/2$ acting on four {\it artificial}
sites, i.e.
\begin{eqnarray}
\sigma_{1}^{\alpha}\!=\!\sigma^{\alpha}\!\otimes1_{2}\otimes1_{2}\otimes1_{2}, & \quad & \sigma_{2}^{\alpha}\!=\!1_{2}\otimes\sigma^{\alpha}\!\otimes1_{2}\otimes1_{2},\nonumber \\
\sigma_{3}^{\alpha}\!=\!1_{2}\otimes1_{2}\otimes\sigma^{\alpha}\!\otimes1_{2}, & \quad & \sigma_{4}^{\alpha}\!=\!1_{2}\otimes1_{2}\otimes1_{2}\otimes\sigma^{\alpha}\!.\label{eq:paul4}
\end{eqnarray}
The Hamiltonian in the gauge transformed basis for the zig-zag with
$L_{z}=3$ can be represented as
\begin{eqnarray}
\tilde{{\cal H}}_{\vec{k}} & \!=\! & \frac{1}{2}\sin\frac{k_{2}}{4}\left(\sigma_{4}^{y}\!-\!\sigma_{3}^{x}\sigma_{4}^{y}\!-\!\sigma_{1}^{z}\sigma_{3}^{y}\sigma_{4}^{x}\!-\!\sigma_{1}^{z}\sigma_{3}^{z}\sigma_{4}^{y}\right)\nonumber \\
 & \!+\! & \frac{1}{2}\cos\frac{k_{2}}{4}\left(-\!\sigma_{4}^{x}\!-\!\sigma_{3}^{x}\sigma_{4}^{x}\!+\!\sigma_{1}^{z}\sigma_{3}^{y}\sigma_{4}^{y}\!+\!\sigma_{1}^{z}\sigma_{3}^{z}\sigma_{4}^{x}\right)\nonumber \\
 & \!+\! & \frac{1}{2}\sin\frac{k_{2}}{4}\sin\frac{k_{1}}{2}\sigma_{2}^{x}\left(\sigma_{3}^{y}\sigma_{4}^{y}\!+\!\sigma_{3}^{z}\sigma_{4}^{x}\!+\!\sigma_{1}^{z}\sigma_{4}^{x}\!+\!\sigma_{1}^{z}\sigma_{3}^{x}\sigma_{4}^{x}\right)\nonumber \\
 & \!+\! & \frac{1}{2}\cos\frac{k_{2}}{4}\sin\frac{k_{1}}{2}\sigma_{2}^{x}\left(\sigma_{3}^{y}\sigma_{4}^{x}\!+\!\sigma_{3}^{z}\sigma_{4}^{y}\!+\!\sigma_{1}^{z}\sigma_{4}^{y}\!-\!\sigma_{1}^{z}\sigma_{3}^{x}\sigma_{4}^{y}\right)\nonumber \\
 & \!+\! & \frac{1}{2}\sin\frac{k_{2}}{4}\cos\frac{k_{1}}{2}\sigma_{1}^{z}\sigma_{2}^{x}\left(\sigma_{3}^{y}\sigma_{4}^{x}\!+\!\sigma_{3}^{z}\sigma_{4}^{y}\!+\!\sigma_{1}^{z}\sigma_{4}^{y}\!-\!\sigma_{1}^{z}\sigma_{3}^{x}\sigma_{4}^{y}\right)\nonumber \\
 & \!-\! & \frac{1}{2}\cos\frac{k_{2}}{4}\cos\frac{k_{1}}{2}\sigma_{1}^{z}\sigma_{2}^{x}\left(\sigma_{3}^{y}\sigma_{4}^{y}\!+\!\sigma_{3}^{z}\sigma_{4}^{x}\!+\!\sigma_{1}^{z}\sigma_{4}^{x}\!+\!\sigma_{1}^{z}\sigma_{3}^{x}\sigma_{4}^{x}\right)\nonumber \\
 & \!+\! & J_{H}\sigma_{2}^{z}-\lambda\sigma_{1}^{y}\sigma_{2}^{z}.
\end{eqnarray}
The spatial symmetries take the form,
\begin{eqnarray}
{\cal \tilde{R}} & = & \frac{1}{2}\sigma_{1}^{y}\left(\left(1-\sigma_{3}^{x}\right)\left(1-\sigma_{4}^{z}\right)-2\right),\nonumber \\
{\cal \tilde{R}}^{t} & = & \sigma_{1}^{y}\sigma_{3}^{x},\nonumber \\
\tilde{{\cal I}} & = & \frac{1}{2}\left(\left(1-\sigma_{3}^{x}\right)\left(1+\sigma_{4}^{z}\right)-2\right),
\end{eqnarray}
and the non-spatial symmetries are
\begin{eqnarray}
{\cal \tilde{T}} & = & \sigma_{1}^{z},\nonumber \\
{\cal \tilde{S}} & = & \sigma_{2}^{x}\sigma_{4}^{z},\nonumber \\
\tilde{C} & = & \sigma_{1}^{z}\sigma_{2}^{x}\sigma_{4}^{z},
\end{eqnarray}
while the shift operators can be expressed as
\begin{eqnarray}
\chi_{1} & = & \sigma_{2}^{z},\nonumber \\
\chi_{2} & = & \frac{1}{\sqrt{2}}\sigma_{3}^{z}\left(i+\sigma_{4}^{z}\right).
\end{eqnarray}
Note that, as expected for $L_{z}=3$, $(\chi_{2})^{4}=1$ whereas
the lower powers are non-trivial, e.g., $(\chi_{2})^{2}=\sigma_{4}^{z}$. 
Finally the second and third neighbor hopping $\delta$ is this basis
has a form,
\begin{equation}
\tilde{h}(\delta)=\delta\sigma_{2}^{z}\sigma_{3}^{x}\cos\frac{k_{2}}{2}.
\end{equation}
One can easily check that it preserves all the symmetries.

\subsection{Symmetry in the parameters space\label{sec:super}}

For any zig-zag segment length $L_{z}$ it is possible to find additional
symmetries that can be associated to two reflections operators $\tilde{{\cal X}}$
and $\tilde{{\cal Y}}$ and an inversion $\tilde{{\cal Z}}=\tilde{{\cal X}}\tilde{{\cal Y}}$
that act uniquely in the parameters space $(J_{H},\lambda)$. The reflections
are active only in the glide plane $k_{1}=0$ and satisfy the relations,
\begin{eqnarray}
\tilde{{\cal X}}{}^{\dagger}\tilde{{\cal H}}_{0,k_{2}}\left(J_{H},\lambda\right)\tilde{{\cal X}} & = & \tilde{{\cal H}}_{0,k_{2}}\left(\lambda,J_{H}\right),\nonumber \\
\tilde{{\cal Y}}{}^{\dagger}\tilde{{\cal H}}_{0,k_{2}}\left(J_{H},\lambda\right)\tilde{{\cal Y}} & = & \tilde{{\cal H}}_{0,k_{2}}\left(-\lambda,-J_{H}\right).
\end{eqnarray}
Thus we see that the reflection planes in the parameters plane $(J_{H},\lambda)$
are in the direction $J_{H}=\lambda$ and $J_{H}=-\lambda$. The action
of the inversion $\tilde{{\cal Z}}$ on the Hamiltonian is obviously,
\begin{equation}
\tilde{{\cal Z}}{}^{\dagger}\tilde{{\cal H}}_{0,k_{2}}\left(J_{H},\lambda\right)\tilde{{\cal Z}}=\tilde{{\cal H}}_{0,k_{2}}\left(-J_{H},-\lambda\right).
\end{equation}
The general matrix form of the two reflections in the gauged basis
is, 
\begin{equation}
\tilde{{\cal X}}=\frac{1}{2}\begin{pmatrix}{\bf 1}_{N_{\downarrow}} & -{\bf 1}_{N_{\downarrow}} & i{\bf 1}_{N_{\downarrow}} & i{\bf 1}_{N_{\downarrow}}\\
-{\bf 1}_{N_{\downarrow}} & {\bf {\bf 1}}_{N_{\downarrow}} & i{\bf 1}_{N_{\downarrow}} & i{\bf {\bf 1}}_{N_{\downarrow}}\\
-i{\bf 1}_{N_{\downarrow}} & -i{\bf 1}_{N_{\downarrow}} & {\bf 1}_{N_{\downarrow}} & -{\bf 1}_{N_{\downarrow}}\\
-i{\bf 1}_{N_{\downarrow}} & -i{\bf {\bf 1}}_{N_{\downarrow}} & -{\bf 1}_{N_{\downarrow}} & {\bf {\bf 1}}_{N_{\downarrow}}
\end{pmatrix},
\end{equation}
and
\begin{equation}
\tilde{{\cal Y}}=-\frac{1}{2}\begin{pmatrix}-{\bf 1}_{N_{\downarrow}} & {\bf 1}_{N_{\downarrow}} & i{\bf 1}_{N_{\downarrow}} & i{\bf 1}_{N_{\downarrow}}\\
{\bf 1}_{N_{\downarrow}} & -{\bf {\bf 1}}_{N_{\downarrow}} & i{\bf 1}_{N_{\downarrow}} & i{\bf {\bf 1}}_{N_{\downarrow}}\\
-i{\bf 1}_{N_{\downarrow}} & -i{\bf 1}_{N_{\downarrow}} & -{\bf 1}_{N_{\downarrow}} & {\bf 1}_{N_{\downarrow}}\\
-i{\bf 1}_{N_{\downarrow}} & -i{\bf {\bf 1}}_{N_{\downarrow}} & {\bf 1}_{N_{\downarrow}} & -{\bf {\bf 1}}_{N_{\downarrow}}
\end{pmatrix}.
\end{equation}
Note that the spectra of $\tilde{{\cal X}}$ and $\tilde{{\cal Y}}$
are the same and consist of $N_{\downarrow}$ eigenvalues $-1$ and
$3N_{\downarrow}$eigenvalues $1$ which concides with the spectrum
of the SU$(2)$ spin interchange operator $X_{12}=\frac{1}{2}(1+\vec{\sigma}_{1}\vec{\sigma}_{2})$
taken $N_{\downarrow}$ times. The spectrum of $\tilde{{\cal Z}}$
consist of eqaul number of $+1$ and $-1$ eigenvalues.

For $L_{z}=2$ and $L_{z}=3$ operators $\tilde{{\cal X}}$, $\tilde{{\cal Y}}$
and ${\cal \tilde{Z}}$ can be written in terms Pauli matrices (\ref{eq:paul3})
or (\ref{eq:paul4}) as,

\begin{eqnarray}
\tilde{{\cal X}} & = & \frac{1}{2}\left(1-\sigma_{1}^{y}-\sigma_{1}^{y}\sigma_{2}^{x}-\sigma_{2}^{x}\right),\nonumber \\
\tilde{{\cal {\cal Y}}} & = & -\sigma_{2}^{x}\tilde{{\cal X}},\nonumber \\
\tilde{{\cal Z}} & = & -\sigma_{2}^{x}.
\end{eqnarray}
For these two shortest zig-zags with $L_{z}=2,3$ we find that the
inversion operator $\tilde{{\cal Z}}$ satisfies,
\begin{equation}
\tilde{{\cal Z}}{}^{\dagger}\tilde{{\cal H}}_{k_{1},k_{2}}\left(J_{H},\lambda\right)\tilde{{\cal Z}}=\tilde{{\cal H}}_{k_{1},k_{2}}\left(-J_{H},-\lambda\right),
\end{equation}
for any $k$-point whereas only for $L_{z}=2$ we get extra relations
for the $\tilde{{\cal X}}$ and $\tilde{{\cal Y}}$ operators in the plane $k_{2}=0$, 
\begin{eqnarray}
\tilde{{\cal X}}{}^{\dagger}\tilde{{\cal H}}_{k_{1},0}\left(J_{H},\lambda\right)\tilde{{\cal X}} & = & \tilde{{\cal H}}_{k_{1},0}\left(\lambda,J_{H}\right),\nonumber \\
\tilde{{\cal Y}}{}^{\dagger}\tilde{{\cal H}}_{k_{1},0}\left(J_{H},\lambda\right)\tilde{{\cal Y}} & = & \tilde{{\cal H}}_{k_{1},0}\left(-\lambda,-J_{H}\right).
\end{eqnarray}

\subsection{Multiple glide and reflection operators\label{sec:mult_refl}}

An interesting consequence of gauge transformation described in Sec.
\ref{sec:hams=000026syms_gaug}, related with elongation of Hamiltonian's
period, is multiplication of reflection and glide operators. Have
a look at the reflection planes, we easily find that for any $L_{z}$,

\begin{equation}
\left[\tilde{{\cal H}}_{k_{1},0},\tilde{{\cal R}}\right]=0,
\end{equation}
but due to the period elongation in $\tilde{{\cal H}}_{k_{1},k_{2}}$
we also find that 
\begin{equation}
\left[\tilde{{\cal H}}_{k_{1},\pi},\tilde{{\cal R}}\right]\not=0.
\end{equation}
This means that $k_{2}=0$ is the reflection plane for $\tilde{{\cal R}}$
but $k_{2}=\pi$ is not. It is not difficult to guess that the sceond
reflection plane should be placed at $k_{2}$ equal to half-period
of the new BZ, namely at $k_{2}=N_{\downarrow}\pi$. Indee we find
\begin{equation}
\left[\tilde{{\cal H}}_{k_{1},N_{\downarrow}\pi},\tilde{{\cal R}}\right]=0,
\end{equation}
but one may ask what about $k_{2}=\pi,2\pi,\dots(N_{\downarrow}-1)\pi$
, is there any reflection operator for whom these are the reflection
planes? The answer is yes, we can define {\it shifted} reflection
operators ${\cal \tilde{R}}_{\chi}^{(n)}$ in the following way,

\begin{equation}
{\cal \tilde{R}}_{\chi}^{(n)}\equiv{\cal \tilde{R}}\left(\chi_{2}\right)^{n},\label{eq:shrefl}
\end{equation}
where $n=1,2,\dots,(N_{\downarrow}-1)$. Their action on Hamiltonian
is the following, 
\begin{equation}
{\cal \tilde{R}}_{\chi}^{(n)\dagger}\tilde{{\cal H}}_{k_{1},k_{2}}{\cal \tilde{R}}_{\chi}^{(n)}=\tilde{{\cal H}}_{k_{1},-k_{2}+2\pi n}.
\end{equation}
Now it's easy to notice that, 
\begin{equation}
\left[\tilde{{\cal H}}_{k_{1},n\pi},{\cal \tilde{R}}_{\chi}^{(n)}\right]=\left[\tilde{{\cal H}}_{k_{1},n\pi+N_{\downarrow}\pi},{\cal \tilde{R}}_{\chi}^{(n)}\right]=0,
\end{equation}
meaning that planes $k_{2}=n\pi$ and $k_{2}=n\pi+N_{\downarrow}\pi$
are the reflection planes for the shifted reflection operator ${\cal \tilde{R}}_{\chi}^{(n)}$.
Note that unlike initial reflection $\tilde{{\cal R}}$ the shifted
reflection operators are not hermitian and unitary, they are only
unitary. Similarly we can define a shifted glide oprator ${\cal \tilde{R}}_{\chi}^{t}$,
\begin{equation}
{\cal \tilde{R}}_{\chi}^{t}\equiv{\cal \tilde{R}}^{t}\chi_{1}.\label{eq:shglide}
\end{equation}
Here we have only one shifted operator because for any zig-zag $(\chi_{1})^{2}=1$.
For this operator the reflection planes are $k_{1}=\pi$ and $k_{1}=3\pi$
whereas for non-shifted ${\cal \tilde{R}}^{t}$ these are $k_{1}=0$
and $k_{1}=2\pi$. Note that the period of gauged Hamiltonian $\tilde{{\cal H}}_{k_{1},k_{2}}$
in $k_{1}$ is $4\pi$ for any $L_{z}$. By taking products of shifted
reflection and glide operators we can construct different shifted
inversion operators for different inversion points in the enlarged
BZ of $\tilde{{\cal H}}_{k_{1},k_{2}}$.

The final conclusion for this Section is that however the $k$-dependence
in spatial symmetry operators can be removed by a proper gauge transformation,
this dependence reappears in the gauged basis as a multiple definition
of these operators for different symmetry-invariant $k$-points.

\section{Reducing the Hamiltonian into a purely real matrix form \label{sec:re_ham}}

The combination of the time reversal and inversion transformation leads to the
operator ${\cal K}_{\vec{k}}$ defined as,
\begin{equation}
{\cal K}_{\vec{k}}\equiv{\cal I}_{\vec{k}}{\cal T},
\end{equation}
whose action on the Hamiltonian is to make it transposed or complex conjugated, 
\begin{equation}
{\cal K}_{\vec{k}}^{\dagger}{\cal H}_{\vec{k}}{\cal K}_{\vec{k}}={\cal H}_{\vec{k}}^{T}.
\end{equation}
Thus, we can generally indicate ${\cal K}_{\vec{k}}$ as a {\it conjugation} operator.
Due to its structure and on the property of $\cal I$ and $\cal T$, we find that the square of ${\cal K}_{\vec{k}}$ gives the identity, i.e., 
\begin{equation}
{\cal K}_{\vec{k}}{\cal K}_{\vec{k}}^{\star}\equiv1 \,.
\end{equation}
From the fact that ${\cal K}_{\vec{k}}$ is unitary, we also deduce that
${\cal K}_{\vec{k}}=\exp(iK_{\vec{k}})$, where
$K_{\vec{k}}$ is a hermitian matrix. Thus, if ${\cal K}_{\vec{k}}{\cal K}_{\vec{k}}^{\star}\equiv1$
then $K_{\vec{k}}$ must be also symmetric and real. 
On this basis, $K_{\vec{k}}$ can be diagonalized by a real unitary transformation
and accordingly for ${\cal K}_{\vec{k}}$. The eigenvalues of ${\cal K}_{\vec{k}}$
are $\pm1$, hence it can be put in a diagonal form ${\cal D}_{{\cal K}}$ by a suitable unitary and {\it real} transformation ${\cal U}_{\vec{k}}$:
\begin{equation}
{\cal D}_{{\cal K}}\equiv{\cal U}_{\vec{k}}^{\dagger}{\cal K}_{\vec{k}}{\cal U}_{\vec{k}}=\begin{pmatrix}{\bf 1} & {\bf 0}\\
{\bf 0} & -{\bf 1} \,.
\end{pmatrix},
\end{equation}

Furthermore, we can introduce another unitary transformation ${\cal V}_{\vec{k}}$
as
\begin{equation}
{\cal V}_{\vec{k}}\equiv{\cal U}_{\vec{k}}\sqrt{{\cal D}_{{\cal K}}},
\end{equation}
in such a way that we can transform the Hamiltonian
to get ${\cal H}'_{\vec{k}}$ in the following form
\begin{equation}
{\cal H}'_{\vec{k}}={\cal V}_{\vec{k}}^{\dagger}{\cal \tilde{{\cal H}}}_{\vec{k}}{\cal V}_{\vec{k}}.
\end{equation}
It is then important to notice that the transformed Hamiltonian is
purely real. To achieve this result one needs to transform $\tilde{{\cal K}}$
by means of $\tilde{{\cal V}}$, recalling that $\tilde{{\cal K}}$
transforms as an anti-unitary operator. Indeed, one obtains
\begin{equation}
{\cal K}'_{\vec{k}}={\cal V}_{\vec{k}}^{T}{\cal K}_{\vec{k}}{\cal V}_{\vec{k}},
\end{equation}
and from the definition of ${\cal V}_{\vec{k}}$ and from the
fact that ${\cal U}_{\vec{k}}$ is real we get 
\begin{equation}
{\cal K}'_{\vec{k}}=1.
\end{equation}
On the other hand, since we know that ${\cal K}'$ satisfies a relation
with ${\cal H}'_{\vec{k}}$ which is given 
\begin{equation}
({\cal K}'_{\vec{k}})^{\dagger}{\cal H}'_{\vec{k}}{\cal K}'_{\vec{k}}=({\cal H}'_{\vec{k}})^{T}.
\end{equation}
we finally conclude that ${\cal H}'_{\vec{k}}\equiv({\cal H}'_{\vec{k}})^{T}$
that implies ${\cal H}'_{\vec{k}}$ to be purely real. 

\section{Reducing the Hamiltonian into a purely imaginary form for odd $L_{z}$\label{sec:imag_ham}}

Combining the pure conjugation and the chirality operators we can
obtain something that we will call an {\it anticonjugation} operator
${\cal A}_{\vec{k}}$, i.e.,
\begin{equation}
{\cal A}\propto{\cal S}_{k_1}{\cal K}_{\vec{k}}.
\end{equation}
Its action on the Hamiltonian is easy to predict, namely,

\begin{equation}
{\cal A}_{\vec{k}}{}^{\dagger}{\cal H}{}_{\vec{k}}{\cal A}_{\vec{k}}=-({\cal H}{}_{\vec{k}})^{T}.
\end{equation}
For zig-zag $L_{z}=2$ ${\cal A}$ turns out to be imaginary 
whereas for $L_{z}=3$ (and any other odd $L_z$) the anticonjugation operator is purely real
(in the gauged basis these are $\tilde{{\cal A}}\equiv\sigma_{1}^{z}\sigma_{2}^{x}\sigma_{3}^{y}$
and $\tilde{{\cal A}}\equiv\frac{1}{2}\sigma_{1}^{z}\sigma_{2}^{x}\left(\left(1-\sigma_{3}^{x}\right)\left(1+\sigma_{4}^{z}\right)-2\sigma_{4}^{z}\right)$
for $L_{z}=2$ and $L_{z}=3$ respectively)
so by analogical unitary transformation to the one described in previous
Section we can obtain a Hamiltonian that satisfies ${\cal H}'_{\vec{k}}\equiv-({\cal H}'_{\vec{k}})^{T}$.
This means that for $L_{z}=3$ we can find a basis where the Hamiltonian
is purely imaginary. Note that this is not possible for $L_{z}=2$
because ${\cal A}$ is imaginary. This difference follows
from the fact that for every even $L_{z}$ the inversion operator
anticommutes with the chirality whereas for odd $L_{z}$ it commutes.

\section{Shift-equivalence of the glide blocks of the Hamiltonian\label{sec:shift_glide}}

In this section we will focus on Hamiltonian in its glide planes.
From Section \ref{sec:mult_refl} we know that in the gauged basis
we should cosider two glide operators, ${\cal \tilde{R}}^{t}$ and
the shifted one ${\cal \tilde{R}_{\chi}}^{t}$, for glide planes $k_{1}=0$
and $k_{1}=\pi$. Looking into Sec. \ref{sec:hams=000026syms_gaug}
we see that both for $L_{z}=2$ and $L_{z}=3$ these operators anticommute
with shift operators $\chi_{2}$, i.e.,

\begin{equation}
\left\{ {\cal \tilde{R}}^{t},\chi_{2}\right\} =0,\quad\left\{ {\cal \tilde{R}_{\chi}}^{t},\chi_{2}\right\} =0,
\end{equation}
and the same property holds for any other $L_{z}$. Now, take the
eigenbasis ${\cal V}$ of ${\cal \tilde{R}}^{t}$ or ${\cal \tilde{R}_{\chi}}^{t}$
and write the two glide plane Hamiltonians as, 
\begin{equation}
\tilde{{\cal H}}'_{0(\pi),k_{2}}\!\equiv\!\tilde{{\cal V}}^{\dagger}\tilde{{\cal H}}_{0(\pi),k_{2}}\!\tilde{{\cal V}}\!=\!\begin{pmatrix}\tilde{H}_{0(\pi),k_{2}}^{+} & 0\\
0 & \tilde{H}_{0(\pi),k_{2}}^{-}
\end{pmatrix},
\end{equation}
where $\tilde{H}_{0(\pi),k_{2}}^{\pm}$ denote the blocks of equal
size in the subspaces of $+1$ and $-1$ eigenvalues of ${\cal \tilde{R}}^{t}$
or ${\cal \tilde{R}_{\chi}}^{t}$ operators. In the same basis we
find $\chi_{2}$ in the antidiagonal form of, 
\begin{equation}
\chi'_{2}\equiv{\cal V}^{\dagger}\chi_{2}{\cal V}=\begin{pmatrix}0 & \xi_{2}\\
\xi_{2} & 0
\end{pmatrix},
\end{equation}
where $\xi_{2}$ is a unitary matrix. From the relation of shift operator
with respect to the Hamiltonian, namely, 
\begin{equation}
\chi'_{2}{}^{\dagger}\tilde{{\cal H}}'_{0(\pi),k_{2}}\chi'_{2}=\tilde{{\cal H}}'_{0(\pi),k_{2}+2\pi},
\end{equation}
we find that, 
\begin{equation}
\xi_{2}^{\dagger}\tilde{H}_{0(\pi),k_{2}}^{\mp}\xi_{2}=\tilde{H}_{0(\pi),k_{2}+2\pi}^{\pm}.
\end{equation}
This a major result that shows that the glide plane Hamiltonian for
$+1$ gliding eigenstates at quasimomentum $k_{2}$ is related with
the Hamiltonian for $-1$ gliding eignestates at point $k_{2}+2\pi$
only by a basis rotation. What more, if we remove the gauging and
come back to original basis we find that these two block are equal,

\begin{equation}
H_{0(\pi),k_{2}}^{\mp}=H_{0(\pi),k_{2}+2\pi}^{\pm}.\label{eq:Hpm_rel}
\end{equation}
Note that $2\pi$ shift in $k_{2}$ is relevant from the point of
view of $H_{0(\pi),k_{2}}^{\mp}$ because now we are in the eigenbasis
of ${\cal R}_{k_{1},k_{2}}^{t}$ which is $k$-dependent and the period
of $H_{0(\pi),k_{2}}^{\mp}$ is elongated. This property, that holds
for any zig-zag segment length $L_{z}$, implies that the whole spectrum
of the glide-plane Hamiltonian is fully determined in just one eigen-subspace
of the glide operator.

\section{Determinant equivalence of the glide blocks \label{sec:det_eq}} 

\subsection{Half-filling case}

In the previous Section we showed that the glide blocks
of the Hamiltonian are related by a $2\pi$ shift as
\begin{equation}
H_{0(\pi),k_{2}}^{\mp}=H_{0(\pi),k_{2}+2\pi}^{\pm}.
\end{equation}
Now we will show that at {\it the same} $k_{2}$ the spectra of $H_{0(\pi),k_{2}}^{\mp}$
are related in a very special way. Namely, for any $L_{z}$ and $k_{1}=0$
we find that,

\begin{equation}
\det H_{0,k_{2}}^{+}\equiv\det H_{0,k_{2}}^{-}\equiv\det H_{0,k_{2}+2\pi}^{+},\label{eq:det_0}
\end{equation}
and for any {\it odd} $L_{z}$ and $k_{1}=\pi$ we have,
\begin{equation}
\det H_{\pi,k_{2}}^{+}\equiv\det H_{\pi,k_{2}}^{-}\equiv\det H_{\pi,k_{2}+2\pi}^{+}.\label{eq:det_pi}
\end{equation}
This means that in each block the product of all eigenvalues
is $2\pi$ periodic although these eigenvalues by themselves have longer period.

Let us show why such property of determinant holds by considering
zig-zag patterns with $L_{z}=3$ and $k_1=0$. We find the determinant of
a glide block to be,
\begin{equation}
\det H_{0,k_{2}}^{+}=\left(-2+\left(J_{H}^{2}-\lambda^{2}\right)^{2}+2\cos k_{2}\right)^{2},
\end{equation}
although the periodicity of $H_{0,k_{2}}^{+}$ is $4\pi$.

Such relation is related to a sort of {\it hidden} symmetry. 
Indeed, the block can be written in the following way
\begin{equation}
H_{0,k_{2}}^{+}=h_{k_{2}}+\bar{H}{}_{k_{2}},
\end{equation}
where $\bar{H}{}_{k_{2}}$ is the $2\pi$ periodic part of $H_{0,k_{2}}^{+}$,
i.e., $\bar{H}{}_{k_{2}+2\pi}\equiv\bar{H}{}_{k_{2}}$ and $h_{k_{2}}$
is the part with $4\pi$ period. Since the dependence on $k_{2}$
is always enclosed in sine and cosine type functions we have
\begin{equation}
h_{k_{2}+2\pi}\equiv-h_{k_{2}}.
\end{equation}
Now, we can write the desired determinant in the following way, 
\begin{equation}
\det H_{0,k_{2}}^{+}\!=\!\det\left(h_{k_{2}}\!+\!\bar{H}{}_{k_{2}}\right)\!=\!\det h_{k_{2}}\det\left(1\!+\!\Sigma_{k_{2}}\right),
\end{equation}
where $\Sigma_{k_{2}}\equiv h_{k_{2}}^{-1}\bar{H}{}_{k_{2}}$. This step requires 
$h_{k_{2}}$ to be an invertible function, and it holds 
because $h_{k_{2}}$ has eigenvalues $\pm\lambda$ so
it is non-singular for any $k_{2}$. The new operator $\Sigma_{k_{2}}$
satisfies the relation $\Sigma_{k_{2}+2\pi}\equiv-\Sigma_{k_{2}}$. It is non-hermitian
and in principle can be non-diagonalizable. 
In our case, we find $\Sigma_{k_{2}}$ to be diagonalizable and its
spectrum to be chiral with the following eigenvalues 
\begin{equation}
s_{k_{2}}=\pm\frac{1}{\lambda}\sqrt{J_{H}^{2}\pm2\sin\frac{k_{2}}{2}},
\end{equation}
and being double degenerate. This means that there exist a non-singular
operator $\beta_{k_{2}}$ that anticommutes with $\Sigma_{k_{2}}$
namely
\begin{equation}
\left\{ \beta_{k_{2}},\Sigma_{k_{2}}\right\} \equiv0.
\end{equation}
Using this property we can prove that determinant of $H_{0,k_{2}}^{+}$
is $2\pi$ periodic as
\begin{equation}
\det H_{0,k_{2}+2\pi}^{+}\!=\!\left(-1\right){}^{N_{\downarrow}}\det h_{k_{2}}\det\left(1\!-\!\Sigma_{k_{2}}\right).
\end{equation}
Then, we focus on the second term and the anticommutation of $\beta_{k_{2}}$,
\[
\det\left(1\!-\!\beta_{k_{2}}^{-1}\beta_{k_{2}}\Sigma_{k_{2}}\right)=\det\left(1\!+\!\beta_{k_{2}}^{-1}\Sigma_{k_{2}}\beta_{k_{2}}\right),
\]
as well as we take into account the Silvester identity that sets the relation between the determinants of two generic matrices $A$ and $B$ 
\begin{equation}
\det\left(1+AB\right)=\det\left(1+BA\right).\label{eq:silv_id}
\end{equation}
Choosing $A=\beta_{k_{2}}^{-1}\Sigma_{k_{2}}$ and $B=\beta_{k_{2}}$
we finally have that,
\begin{equation}
\det\left(1\!+\!\beta_{k_{2}}^{-1}\Sigma_{k_{2}}\beta_{k_{2}}\right)=\det\left(1\!+\!\Sigma_{k_{2}}\right),
\end{equation}
and thus
\begin{equation}
\det\left(1\!-\!\Sigma_{k_{2}}\right)=\det\left(1\!+\!\Sigma_{k_{2}}\right).
\end{equation}
This implies that
\begin{equation}
\det H_{0,k_{2}+2\pi}^{+}\!=\!\left(-1\right){}^{N_{\downarrow}}\det H_{0,k_{2}}^{+}.
\end{equation}
Since $N_{\downarrow}$ is always even in our zig-zag patterns we 
succeed in demonstrating that the determinant of the glide block is indeed $2\pi$
periodic. We point out that a crucial ingredient for the proof
is given by the existence of an {\it invertible} operator $\beta_{k_{2}}$ that
anticommutes with $\Sigma_{k_{2}}$. We found that such chirality
also occurs for $k=\pi$ glide plane and for other zig-zag segment
lengths $L_{z}$. 

\subsection{Away from half-filling}


The property of determinat of a glide block described in the previous
Section can be more general in case of some $L_{z}$. Namely, we can
find such values of chemical potential $\mu$ that a following relation
is satisfied, 
\begin{equation}
\det\left(H_{0,k_{2}}^{+}-\mu\right)\equiv\det\left(H_{0,k_{2}+2\pi}^{+}-\mu\right).\label{eq:det_mu}
\end{equation}
In case of zig-zag $L_{z}=3$ we find that there is one non-trivial
value of $\mu$ satisfying this relation, 

\begin{equation}
\mu_{0}=\pm\sqrt{2+J_{H}^{2}+\lambda^{2}},
\end{equation}
where the freedom of sign comes from the chirality of $H_{0,k_{2}}^{+}$.
The determinant then becomes,

\begin{equation}
\det\left(H_{0,k_{2}}^{+}-\mu_{0}\right)=4\left(1-2J_{H}^{2}\lambda^{2}+\cos k_{2}\right)^{2}.
\end{equation}
So indeed it is $2\pi$ periodic. Why it happens we can prove in a
indirect way. We define a chiral-square block $H_{0,k_{2}}^{+(2)}$
as, 
\begin{equation}
H_{\mu,k_{2}}^{+(2)}\equiv(H_{0,k_{2}}^{+})^{2}-\mu^{2}.
\end{equation}
For this block we can prove using the method from the previous section
that,
\begin{equation}
\det H_{\mu,k_{2}}^{+(2)}=\det H_{\mu,k_{2}+2\pi}^{+(2)}.\label{eq:det_id_sq}
\end{equation}
Now having this we can relate the determinant of $H_{\mu,k_{2}}^{+(2)}$
with determinant of $(H_{0,k_{2}}^{+}-\mu)$ (modulo sign) in a following
way, 
\begin{eqnarray}
\det H_{\mu,k_{2}}^{+(2)} & = & \det\left(H_{0,k_{2}}^{+}+\mu\right)\det\left(H_{0,k_{2}}^{+}-\mu\right)\nonumber \\
 & = & \left(-1\right){}^{N_{\downarrow}}\det\left(H_{0,k_{2}}^{+}-\mu\right)^{2},
\end{eqnarray}
where for the second equality we used the chirality operator of $H_{0,k_{2}}^{+}$
and the Silvester identity of Eq. (\ref{eq:silv_id}) in the same
way as we did in previous section for $\Sigma_{k_{2}}$ and $\beta_{k_{2}}$. 

The prove of property (\ref{eq:det_id_sq}) can be done in way described
in the previous Section. First we decompose $H_{\mu,k_{2}}^{+(2)}$,
\begin{equation}
H_{\mu,k_{2}}^{+(2)}=h_{k_{2}}^{(2)}+\bar{H}{}_{\mu,k_{2}}^{(2)},
\end{equation}
into the part which is $2\pi$ periodic - $\bar{H}{}_{\mu,k_{2}}^{(2)}$
and the rest, $h_{k_{2}}^{(2)}$ satisfying $h_{k_{2}+2\pi}^{(2)}\equiv-h_{k_{2}}^{(2)}$.
Now we define the operator $\Sigma_{\mu,k_{2}}^{(2)}$ which we would
like to be chiral, i.e., 
\begin{equation}
\Sigma_{\mu,k_{2}}^{(2)}\equiv h_{k_{2}}^{(2)-1}\bar{H}{}_{\mu,k_{2}}^{(2)}.
\end{equation}
Indeed the spectrum of $\Sigma_{\mu,k_{2}}^{(2)}$ is chiral if only
$\mu=\mu_{0}$ but there is one subtelty here - $\Sigma_{\mu_{0},k_{2}}^{(2)}$
is non-diagonalizable (defective). We find that it has a non-trivial
Jordan form given by,
\begin{equation}
\Sigma_{\mu_{0},k_{2}}^{(2)'}=\begin{pmatrix}-s_{k_{2}} & 1 & 0 & 0 & 0 & 0 & 0 & 0\\
0 & -s_{k_{2}} & 0 & 0 & 0 & 0 & 0 & 0\\
0 & 0 & -s_{k_{2}} & 1 & 0 & 0 & 0 & 0\\
0 & 0 & 0 & -s_{k_{2}} & 0 & 0 & 0 & 0\\
0 & 0 & 0 & 0 & s_{k_{2}} & 0 & 0 & 0\\
0 & 0 & 0 & 0 & 0 & s_{k_{2}} & 0 & 0\\
0 & 0 & 0 & 0 & 0 & 0 & s_{k_{2}} & 0\\
0 & 0 & 0 & 0 & 0 & 0 & 0 & s_{k_{2}}
\end{pmatrix}\!,
\end{equation}
with eigenvalues, 
\begin{equation}
\pm s_{k_{2}}\equiv\pm\frac{1}{J_{H}\lambda}\cos\frac{k_{2}}{2},
\end{equation}
 and where $\Sigma_{\mu_{0},k_{2}}^{(2)'}$ is related with $\Sigma_{\mu_{0},k_{2}}^{(2)}$
by a similarity transformation, 
\begin{equation}
\Sigma_{\mu_{0},k_{2}}^{(2)'}=\gamma^{-1}\Sigma_{\mu_{0},k_{2}}^{(2)}\gamma.
\end{equation}
The fact that $\Sigma_{\mu_{0},k_{2}}^{(2)}$ is defective means that
we cannot find a non-singular matrix $\beta_{\mu_{0},k_{2}}^{(2)}$
that anticommutes with $\Sigma_{\mu_{0},k_{2}}^{(2)}$ eventhough
its spectrum is chiral. This is however not a big complication because
the non-diagonal entries in $\Sigma_{\mu_{0},k_{2}}^{(2)'}$ do not
affect the determinant of $(1+\Sigma_{\mu_{0},k_{2}}^{(2)'})$ which
is important for the proof. Hence we can replace $\Sigma_{\mu_{0},k_{2}}^{(2)}$
by a new operator $\bar{\Sigma}_{\mu_{0},k_{2}}^{(2)}$ whose form
in the basis given by $\gamma$ is purely diagonal and is identical
to $\Sigma_{\mu_{0},k_{2}}^{(2)'}$ without non-diagonal entries.
For this oparator one can find an anticommuting and non-singular partner
and thus the proof is complete.

\section{Topological invariants \label{sec:topo_inv}}

To calculate the topological invariants we use an approach based
on Green's function \cite{Zha13}. Namely, we define the Green's operator
${\cal G}$ as, 
\begin{equation}
{\cal G}(\omega,k)=\frac{1}{i\omega-{\cal H}_{\vec{k}}},
\end{equation}
where the Fermi energy is at $\omega=0$. For the non-chiral
case of a Fermi surface with a codimension
$p$, being the difference of the system's dimension $d$ and that one 
of the Fermi surface $d_{FS}$, the topological number $N_{p}$ can be expressed as an integral over
an oriented manifold of the dimension $p$, e.g. a $p$-sphere, in
a $(\omega,\vec{k})$-space enclosing the Fermi surface, 
\begin{equation}
N_{p}=C_{p}\int_{S^{p}}{\bf tr}\left[\left({\cal G}{\rm d}{\cal G}^{-1}\right)^{p}\right],\label{eq:Np}
\end{equation}
where the prefactor $C_{p}$ is given by 
\begin{equation}
C_{p}=-\frac{n!}{(2n+1)!(2\pi i)^{n+1}},
\end{equation}
with $p=2n+1$. Thus the formula is valid only for odd $p$ and for
even ones the $\mathbb{Z}$ topological number vanishes. Note that the power
under the trace means an external product of $p$ copies of $({\cal G}{\rm d}{\cal G}^{-1})$.
This formula is used to calculate the topological number (charge) of
a line Fermi surface within the AI class. Because the problem is two-dimensional
we have $p=1$ and thus we can get a non-vanishing $N_{p}$ calculating
the integral over a circle around the Fermi line. For simplicity the
circle can be chosen in the $(\omega,k_{1})$-plane with a center
belonging to the Fermi surface.

In the presence of a chiral symmetry ${\cal S}$, the $\mathbb{Z}$ topological
number lives only at $\omega=0$ so the effective dimension of
the integration is reduced by $1$. Consequently, the chiral $\mathbb{Z}$ topological
number $\nu_{p}$ for the Fermi surface with a codimension $p$ is
defined by, 
\begin{equation}
\nu_{p}=\frac{C_{p-1}}{2}\int_{S^{p-1}}{\bf tr}\left[{\cal S}\left({\cal H}^{-1}{\rm d}{\cal H}\right)^{p-1}\right],\label{eq:nup}
\end{equation}
where the sphere $S^{p-1}$ is only in the $\vec{k}$-space. From
this formula we can calculate the winding numbers of the chiral Dirac
points within the BDI class. 

Finally, we also use the $\mathbb{Z}_{2}$ topological numbers of the first
generation - $\mathbb{Z}_{2}^{(1)}$. These numbers are defined by similar
integrals as the $\mathbb{Z}$-numbers but they require an extension of the
Hamiltonian (or the Green's function). This extension involves an
auxiliary parameter $u\in[0,1]$ which becomes an extra dimension
to integrate over. The extended Hamiltonian has a form ${\cal \tilde{H}}_{\vec{k}}=(1-u){\cal H}_{\vec{k}}+u{\cal H}_{0}$,
where ${\cal H}_{0}$ is a trivial Hamiltonian with energies $\pm E_{0}$.
From the extended Hamiltonian $\tilde{{\cal H}}$ we deduce the 
Green's function $\tilde{{\cal G}}$. The $\mathbb{Z}_{2}^{(1)}$ topological
number $N_{p}^{(1)}$ of the Fermi surface with codimension $p$ is then
given by
\begin{equation}
N_{p}^{(1)}=C_{p}^{'}\int_{S^{p}}\int_{0}^{1}{\rm d}u\:{\bf tr}\left[\left(\tilde{{\cal G}}{\rm d}\tilde{{\cal G}}{}^{-1}\right)^{p}\tilde{{\cal G}}\partial_{u}\tilde{{\cal G}}{}^{-1}\right]\,{\rm mod}\,2,\label{eq:Np(1)}
\end{equation}
with a prefactor,
\begin{equation}
C_{p}^{'}=-\frac{2(p/2)!}{p!(2\pi i)^{p/2+1}}.
\end{equation}
Thus, the $\mathbb{Z}_{2}^{(1)}$- number is non-vanishing only for even codimension
$p$. In our case we use this formula to calculate the topological
charges of the Dirac points in two dimensions- $p=2$ within an effective
D class. It is worth to mention that in this case, when we deal with
a purely real $2\times2$ Hamiltonian, the extension ${\cal H}_{0}$
must be chosen as imaginary to get a non-vanishing $N_{p}^{(1)}$
- here we chose ${\cal H}_{0}=\sigma^{y}$. In the end, we also need
a chiral version of the $N_{p}^{(1)}$ invariant, namely the chiral
$\mathbb{Z}_{2}^{(1)}$- number $\nu_{p}^{(1)}$. This is defined in a usual
way by
\begin{equation}
\nu_{p}^{(1)}\!\!=\!\frac{C_{p-1}^{'}}{2}\!\!\!\int_{\!S^{p-1}}\!\int_{0}^{1}\!\!\!{\rm d}u\:{\bf tr}\!\left[{\cal S}\!\left(\!{\cal \tilde{H}}^{-\!1}\!{\rm d}{\cal \tilde{H}}\right)^{\!p-\!1}\!\!{\cal {\cal \tilde{H}}}^{-\!1}\partial_{u}{\cal \tilde{H}}\right]\,{\rm mod}\,2.\label{eq:nup(1)}
\end{equation}
Such an invariant are used to characterize the Dirac points in the one-dimensional
cut in the BZ.


\begin{thebibliography}{99}
\bibitem{Rem} For sake of completeness, a hermitian congruence is found between the blocks
$H^+_{0,k_2}$ and $H^+_{0,k_2+2\pi}$, thus meaning that in a given basis
these blocks differ only by a scaling of the rows and columns entries. The scaling relation can 
be expressed via four positive 
coefficients which are functions of $k_2$ and their product is equal to $1$.
\end{thebibliography}
\end{document}